\documentclass[a4paper,latin9]{aa}
\usepackage[latin9]{inputenc}
\usepackage{amsmath}

\usepackage{amssymb}
\usepackage{graphicx}
\usepackage{url}

\makeatletter

\providecommand{\tabularnewline}{\\}

\usepackage{times}\usepackage{multirow}\usepackage{rotating}\usepackage{longtable}\usepackage{lscape}
\usepackage{lipsum}
\usepackage{times}
\usepackage{multirow}
\usepackage{rotating}
\usepackage{longtable}
\usepackage{lscape}\usepackage{epstopdf}
\defcitealias{pierre15}{Paper I}
\defcitealias{pacaud15}{Paper II}
\defcitealias{giles15}{Paper III}
\defcitealias{lieu15}{Paper IV}
\defcitealias{pompei15}{Paper VII}
\defcitealias{koulouridis15}{Paper XII}

\def\f#1   {Fig.~\ref{#1}}
\def\s#1   {Sect.~\ref{#1}}
\def\tab#1   {Tab.~\ref{#1}}
\def\eq#1   {Eq.~\ref{#1}}
\def\t#1   {Tab.~\ref{#1}}

\def\comm#1   {{\tt (COMMENT: #1) }}

\titlerunning{GMRT-XXL-N 610 MHz}
\authorrunning{Smolcic et al.}

\makeatother

\begin{document}

\title{The XXL Survey: XXIX. GMRT 610 MHz continuum observations}

\author{
Vernesa~Smol{\v{c}}i{\'{c}}\inst{1}, 
Huib~Intema\inst{2},
Bruno~\v{S}laus\inst{1},
Somak~Raychaudhury$^{3,4}$,
Mladen~Novak\inst{1}, 
Cathy~Horellou$^{5}$,
Lucio~Chiappetti\inst{6}
Jacinta~Delhaize\inst{1}, 
Mark~Birkinshaw\inst{7},
Marco~Bondi$^{8}$,
Malcolm~Bremer$^{7}$, 
Paolo~Ciliegi$^{9}$,
Chiara~Ferrari$^{10}$, 
Konstantinos~Kolokythas$^{3}$,
Chris~Lidman$^{11}$,
Sean~L.~McGee$^{3}$,
Ray~Norris$^{12,13}$,
Marguerite~Pierre$^{14,15}$,
Huub~R\"{o}ttgering$^{2}$,
Cyril~Tasse$^{16}$,
Wendy~Williams$^{17}$
%Huub R\"{o}ttgering$^{2}$,
%Jacinta Delhaize\inst{1}, 
%Mark~Birkinshaw\inst{5},
%Chiara Ferrari$^{6}$, 
%Sean L. McGee$^{3}$,
%Malcolm Bremer$^{5}$, 
%Marguerite Pierre$^{7}$,
%Cyril Tasse$^{8}$,
%Wendy Williams$^{9}$, 
%Konstantinos Kolokythas$^{3}$,
%Ray Norris$^{10,11}$,
%Paolo Ciliegi$^{12}$,
%Marco Bondi$^{13}$
}

\institute{
%1
Department of Physics, Faculty of Science, University of Zagreb,  Bijeni\v{c}ka cesta 32, 10000  Zagreb, Croatia  
\and 
%2
Leiden Observatory, Leiden University, Niels Bohrweg 2, 2333 CA, Leiden, The Netherlands
\and
%3
Inter-University Centre for Astronomy and Astrophysics, Ganeshkhind, Pun\'e-411 007, India
\and
School of Physics and Astronomy, University of Birmingham, Birmingham B15~2TT, UK
%4
\and
%5
Dept. of Space, Earth and Environment, Chalmers University of Technology, Onsala Space Observatory, SE-439 92 Onsala, Sweden
\and
%6
INAF, IASF Milano, via Bassini 15, 20133, Milano, Italy
\and
%7
H.H. Wills Physics Laboratory, University of Bristol, Tyndall Avenue, Bristol BS8 1TL, U.K.
\and
%8
INAF - Istituto di Radioastronomia di Bologna, via P. Gobetti, 101, I-40129, Bologna, Italy 
\and
%9
INAF - Osservatorio  Astronomico di Bologna, via P.Gobetti 93/3, I-40129 Bologna, Italy
\and
%10
Laboratoire Lagrange, Universit\'e C\^ote d'Azur, Observatoire de la C\^ote d'Azur, CNRS,  Blvd de l'Observatoire, CS 34229, 06304 Nice cedex 4, France
  \and
%11
Australian Astronomical Observatory, North Ryde, NSW 2113, Australia
\and
%12
Western Sydney University, Locked Bag 1797, 1797, Penrith South, NSW, Australia
\and
%13
CSIRO Astronomy and Space Science, PO Box 76, 1710, Epping, NSW, Australia
\and
%14
IRFU, CEA, Université Paris-Saclay, F-91191 Gif-sur-Yvette, France %Service d'Astrophysique AIM, CEA/DSM/IRFU/SAp, CEA Saclay, F-91191 Gif sur Yvette 
\and
%15
Université Paris Diderot, AIM, Sorbonne Paris Cité, CEA, CNRS, F-91191
Gif-sur-Yvette, France
\and
%16
GEPI, Observatoire de Paris, CNRS, Universit\'e Paris Diderot, 5 place Jules Janssen, 92190, Meudon, France
\and
%17
School of Physics, Astronomy and Mathematics, University of Hertfordshire, College Lane, Hatfield AL10 9AB, UK
}

\abstract{
We present the 25 square-degree GMRT-XXL-N 610 MHz radio continuum survey, conducted at 50~cm wavelength   with the Giant Metrewave Radio Telescope (GMRT) towards the XXL Northern field (XXL-N). We combined previously published observations of the XMM-Large Scale Structure (XMM-LSS) field, located in the central part of XXL-N, with newly conducted observations towards the remaining XXL-N area, and imaged the combined data-set using the Source Peeling and Atmospheric Modeling ({\sc SPAM}) pipeline. The final mosaic encompasses a total area of $30.4$ square degrees, with ${\rm rms}<150~\mu$Jy/beam over 60\% of the area. The ${\rm rms}$ achieved in the inner 9.6 square degree area, enclosing the XMM-LSS field, is about $200~\mu$Jy/beam, while that over the outer 12.66 square degree area (which excludes the noisy edges) is about $45~\mu$Jy/beam. The resolution of the final mosaic is 6.5 arcsec. We present a catalogue of 5\,434 sources detected at $\geq7\times {\rm rms}$. 
We verify, and correct the reliability of, the catalog in terms of astrometry, flux, and false detection rate. Making use of the (to date) deepest radio continuum survey over a relatively large (2 square degree) field, complete at the flux levels probed by the GMRT-XXL-N survey, we also assess the survey's incompleteness as a function of flux density. The radio continuum sensitivity reached over a large field with a wealth of multi-wavelength data available makes the GMRT-XXL-N 610 MHz survey an important asset for studying the physical properties, environments and cosmic evolution of radio sources, in particular radio-selected active galactic nuclei (AGN).
}

\keywords{Surveys; galaxies: clusters: general, active; radiation mechanisms:
general; radio continuum: galaxies}

\maketitle
\makeatother

% \date{Accepted 20XX Month XX. Received 20XX Month XX; in original form% 20XX Month XX}

%\pagerange{\pageref{firstpage}--\pageref{lastpage}} \pubyear{2011}

%\maketitle

\section{Introduction\label{sec:intro}}

Multiwavelength sky surveys provide a powerful way to study how galaxies and structure form in the early universe and subsequently evolve through cosmic time. These surveys grow both in area and depth with the combined efforts of large consortia and the advent of observational facilities delivering a significant increase in sensitivity.
In this context, the XXL Survey represents the largest XMM-Newton project to date (6.9 Ms; \citealt{pierre16}, hereafter XXL Paper I). It encompasses two areas, each covering 25 square degrees with an X-ray  (0.5-2 keV) point-source sensitivity of $\sim5\times10^{-15}{\rm \ erg\, s^{-1}cm^{-2}}$. A wealth of multiwavelength data (X-ray to radio) is available in both fields. Photometric redshifts are computed for all sources, and %each accuracies better than $\sim10\%$, sufficient for large scale structure studies, and evolutionary studies of galaxies and active galactic nuclei (AGN). 
over $15\,000$ optical spectroscopic redshifts are already available. The main goals of the project are to constrain the dark energy equation of state using clusters of galaxies, and to provide long-lasting legacy data for studies of galaxy clusters and AGN (see XXL Paper I for an overview). 

In the context of AGN and their cosmic evolution, the radio wavelength window offers an important complementary view to X-ray, optical, and infrared observations (e.g. \citealt{padovani17}).  Only via radio observations can AGN hosted by otherwise passive galaxies be revealed (\citealt[e.g.][]{sadler07,smo09,smo17c}), presumably tracing a mode of radiatively inefficient accretion onto the central supermassive black holes, occurring at low Eddington ratios and through puffed-up, geometrically thick but optically thin accretion disks (see \citealt{heckman14} for a review). Furthermore, radio continuum observations directly trace AGN deemed responsible for radio-mode AGN feedback, a key process in semi-analytic models that limits the formation of overly massive galaxies \citep[e.g.][]{croton06,croton16}, a process that still needs to be verified observationally \citep[e.g.][]{smo09a,smo17c,best14}. 

Radio continuum surveys, combined with multiwavelength data are necessary to study the properties of radio AGN at intermediate and high redshifts, their environments , and their cosmic evolution. Optimally,  the sky area surveyed and the sensitivity reached are simultaneously maximised. In practice this is usually achieved through a 'wedding-cake approach' where deep, small area surveys are combined with larger area, but shallow surveys (see e.g. Fig.~1 in \citealt{smo17}). The newly obtained radio continuum coverage of the XXL-North and -Sout fields (XXL-N and XXL-S respectively) is important in this respect, as it covers an area as large as 50 square degrees down to intermediate radio continuum sensitivities (rms~$\sim40-200~\mu$Jy/beam), expected to predominantly probe radio AGN through cosmic time.

The XXL-S has been covered at 843 MHz by the Sydney University Molonglo Sky Survey (SUMSS) down to a sensitivity of 6 mJy/beam \citep{bock99}. To achieve a higher sensitivity, it was observed by the XXL consortium with the Australia Telescope Compact Array (ATCA) at 2.1 GHz  frequency down to a $1\sigma$ sensitivity of $\sim40~\mu$Jy/beam  (\citealt{smolcic16}, hereafter XXL Paper XI; \citealt{butler17}, hereafter XXL Paper XVIII).   

The XXL-N has been covered by the 1.4 GHz (20 cm) NRAO VLA Sky Survey (NVSS; \citealt{condon98}) and Faint Images of the Radio Sky at Twenty Centimeters (FIRST; \citealt{becker95}) surveys down to sensitivities of 0.45 and 0.15 mJy/beam, respectively. Subareas were also covered at 74, 240, 325 and 610 MHz within the XMM-LSS Project (12.66 deg$^2$; \citealt{tasse06, tasse07}), and 610 MHz and 1.4 GHz within the VVDS survey (1 deg$^2$; \citealt{bondi03}). Here we present new GMRT 610 MHz data collected towards the remainder of the XXL-N field. We combine these data with the newly processed 610 MHz data from the XMM-LSS Project, and present a validated source catalogue extracted from the total area observed. This sets the basis necessary for  exploring  the physical properties, environments and cosmic evolution of radio AGN in the XXL-N field (see Horellou et el., XXL Paper XXXIV, in prep.). Combined with the ATCA-XXL-S 2.1 GHz survey (XXL Papers XI and XVIII) it provides a unique radio data set that will allow studies of radio AGN and of their cosmic evolution over the full 50 square degree XXL area. An area of this size is particularly sensitive to probing the rare, intermediate- to high radio-luminosity AGN at various cosmic epochs \citep[e.g.][]{willott01,sadler07,smo09a,donoso09,pracy16}.

The paper is outlined as follows. In \s{sec:data} \ we describe the observations, data reduction, and imaging. The mosaicing procedure and source catalogue extraction are presented in \s{sec:mosaic} , and  \s{sec:catalog} , respectively. We test the reliability and completeness of the catalogue in \s{sec:tests} , and summarise our results in \s{sec:summary} . The radio spectral index, $\alpha$, is defined via $S_\nu\propto\nu^\alpha$, where $S_\nu$ is the flux density at frequency $\nu$.

\section{Observations, data reduction and imaging}
\label{sec:data}

We describe the GMRT 610 MHz observations towards the XXL-N field, and briefly outline the data reduction and imaging performed on this data set using the Source Peeling and Atmospheric Modeling ({\sc SPAM}) pipeline\footnote{\url{http://www.intema.nl/doku.php?id=huibintemaspam}}.

\subsection{Observations}
\label{sec:obs}

Observations with the GMRT at 610 MHz were conducted towards an area of 12.66 deg$^2$ within the XXL-N field (which includes the XMM-LSS area), using a hexagonal grid of 36 pointings (see  \citealt{tasse07} for details). A  total of $\sim18$ hours of observations were taken in the period from July  to  August 2004 \citep{tasse07}\footnote{Project ID 06HRA01}. The full available bandwidth of 32 MHz was used. The band was split into two intermediate frequencies (IFs),  each sampled by
128 channels, and covering the ranges of 594-610 MHz  and 610-626 MHz, respectively. The source 3C 48 was observed for 30 minutes at the beginning and end of an observing run for flux and bandpass calibration, while source 0116-208 was observed for 8 min every 30 min as the secondary calibrator. To optimise the $uv$-coverage \citet{tasse07} split the
30 min observation of each individual pointing into three 10-minute
scans, separated by about 1.3 hours. 

The remaining areas of the XXL-N field, not previously covered at 610 MHz frequency, were observed with the GMRT through Cycles 23\footnote{Project ID 23$\_$022; 30 hours allocated in the period of October 2012 - March 2013}, 
24\footnote{Project ID 24$\_$043; 45 hours allocated in the period of April 2013 - September 2013}, 
27\footnote{Project ID 27$\_$009; 70 hours allocated in the period of October 2014 - March 2015}, and  
30\footnote{Project ID 30$\_$005; 29 hours allocated in the period of April - September 2016} for a total of 174 hours, in a combination of rectangular and hexagonal pointing patterns.  The observations were conducted under good weather conditions. Using the GMRT software backend (GSB)  a total bandwidth of 32 MHz was covered, at a central frequency of 608~MHz, and sampled by 256 channels in total. To maximise the $uv$-coverage, individual pointing scans were spacedout and iterated throughout the observing run whenever possible.
Primary calibrators (3C~48, 3C~147, 3C~286) were observed for an on-source integration time of $10-15$ minutes at the beginning and end of each observing run. Secondary calibrators were also observed multiple times during the observations. We note, however, that phase/amplitude calibration was not performed using the secondary calibrators, but via self-calibration against background models, which has been shown to improve the final output (see \citealt{intema16} for further details; see also below).

%\begin{figure}[h!]
%\includegraphics[width=\columnwidth]{XXL_GMRT_cycle30.eps}
%\protect\protect\caption{
%The pointing layout of GMRT observations performed at 610 MHz towards the XXL-N field.  \label{fig:pts}
%}
%\end{figure}

\subsection{Data reduction and imaging}
\label{sec:spam}

The data reduction and imaging was performed using the {\sc SPAM} pipeline, described in detail by \citet[see also \citealt{intema09, intema14}]{intema16}. The pipeline includes direction dependent calibration, modelling and imaging for correcting mainly ionospheric dispersive delay. It consists of two parts. In the first, pre-processing, step the raw data  from individual observing runs are calibrated using good-quality instrumental calibration solutions obtained per observing run, and then converted into  visibility data sets for each observed pointing. Flagging, gain, and bandpass calibrations are performed in an iterative process, applying increasingly strict radio frequency interference flagging to optimise the calibration results. 
In the second step the main pipeline converts the individual pointing visibility data sets into Stokes I continuum images, performing several steps of direction-independent and direction-dependent calibration, self-calibration, flagging and image construction. Imaging is performed via a single CLEAN deconvolution, automatically setting boxes around sources, and cleaning down to 3 times the central background noise. We refer to \citet{intema16} for further details about the pipeline. %\comm{Huib, what happens in cases a pointing was observed over multiple runs? Are the uv-data combined prior to imaging, or is the combination done in the image plane when mosaicing? What else should we stress here about the pipeline?}

The {\sc SPAM} pipeline successfully processed all XXL-N GMRT 610 MHz observing runs. %\comm{we had a problem with one observing run/pointing, correct?}  
A visual verification of the image quality found satisfactory results for every pointing. %The average rms reached per pointing is about xxx $\mu$Jy/beam in the XMM-LSS area (with an observing time of 30 minutes per pointing), and xxx $\mu$Jy/beam in the remaining XXL-N area (with an average observing time of 3.35 hours per pointing).

\section{Mosaicing}
\label{sec:mosaic}

In this section we describe the astrometric corrections (\s{sec:astrocorr} ) and flux density corrections (i.e., primary beam  and pointing; \s{sec:fluxcorr} ) performed prior to combining the individual pointings into the final mosaic (\s{sec:finalmosaic} ). We constrain and/or verify these corrections using compact sources (total signal-to-noise ratio > 10) 
%a sample of %xxx bright ($S_\mathrm{610MHz}>xxx~\mu$Jy), compact %(major axis $<xx\arcsec$) s
in overlapping pointings,  lying within the inner part of each individual pointing, and extracted with the {\sc PyBDSF}\footnote{\url{http://www.astron.nl/citt/pybdsf/}} software \citep{mohan15}, in the same way as described in \s{sec:pybdsm} . %\comm{one more sentence on details of extraction?}
 A sample assembled in this way assures that the errors on individual flux density/position measurements by Gaussian fitting are minimised so that the noise-independent calibration errors can be determined.

\subsection{Astrometric corrections}
\label{sec:astrocorr}

%\comm{a sentence or two explaining why residual offsets may remain needed} \

To account for possible residual systematic astrometric shifts across individual pointings, initially caused by the ionosphere and not fully accounted for by the direction-dependent calibration, the source positions in each pointing were matched to those in the FIRST survey catalogue (\citealt{becker95}; see also Sect.~\ref{sec:mosaicastrom}), and the systematics corrected accordingly. In total there were 1\,286 sources used for this comparison. %\comm{what are this shifts that we find?}

In \f{fig:astr} \ we show the resulting relative positional offsets of bright, compact sources in overlapping pointings. We find a $1\sigma$ scatter of $0.67\arcsec$ in RA and Dec, with a median offset of $0.02\arcsec$, and  $-0.03\arcsec$ in RA and Dec, respectively, affirming that systematics have been corrected for.

\begin{figure}
\includegraphics[width=\columnwidth]{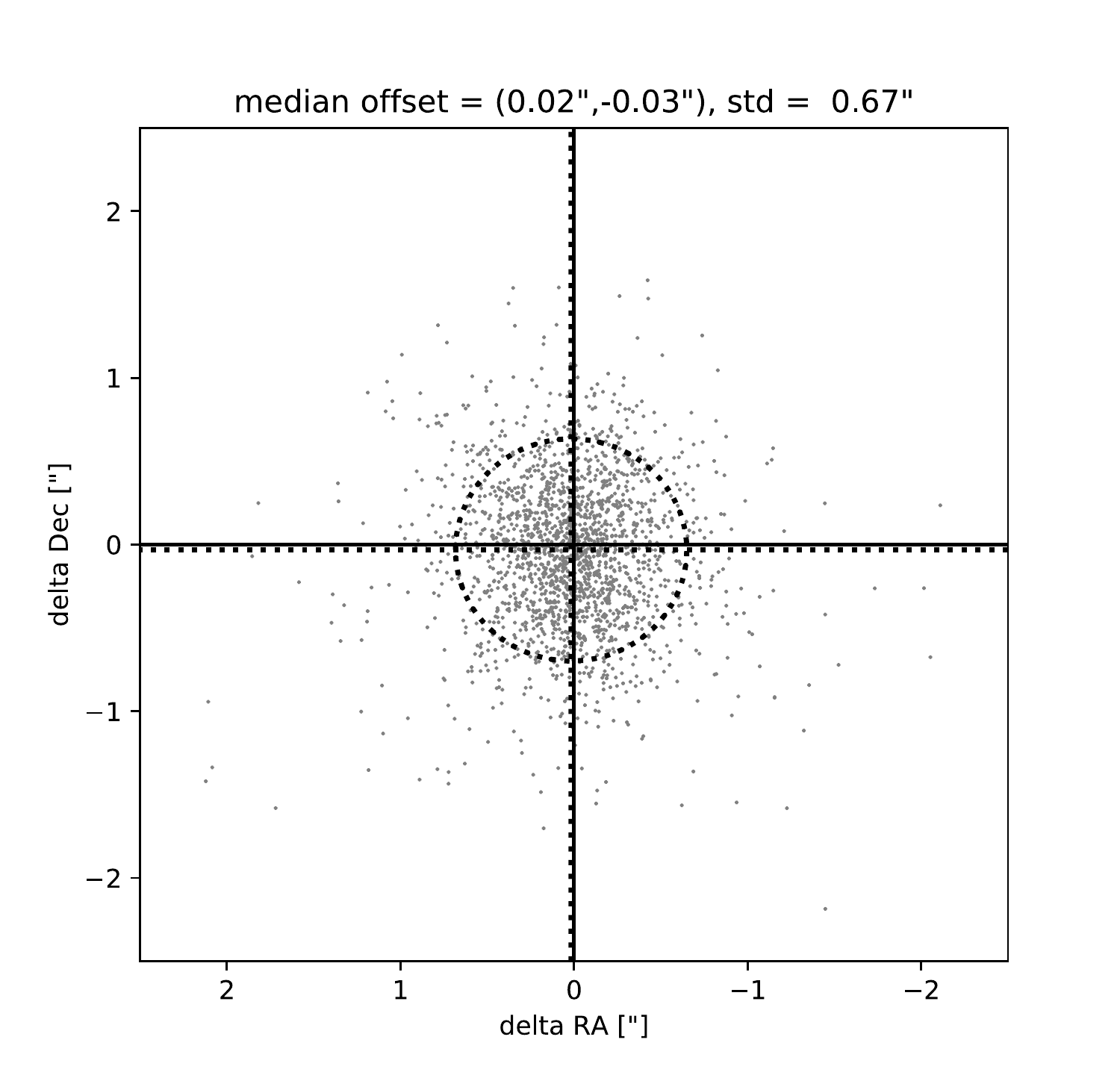}
\protect\protect\caption{
Positional offsets of bright, compact sources in overlapping pointings. The median offsets in RA and Dec are indicated by the vertical and horizontal lines, respectively, while the dotted circle represents the $1\sigma$ deviation (also indicated at the top).
\label{fig:astr}
}
\end{figure}

\subsection{Flux density corrections}
\label{sec:fluxcorr}
%\subsection{Primary beam corrections}
%\label{sec:pbcorr}

To correct the individual pointing maps for primary beam attenuation, 
 we adopt a standard, parameterised axisymmetric model, with coefficients given in the GMRT Observer Manual. Given the small fractional bandwidth covered (5.25\%), we use the central frequency (610~MHz) beam model for all frequency channels.
In \f{fig:pb} \ we show the ratio of the flux densities of the sources in overlapping pointings, but at various distances from the phase centres, not corrected for primary beam attenuation (hereafter apparent flux densities) as a function of the ratio of the primary beam model attenuations
for the same sample of  compact sources in overlapping pointings. For a perfect primary beam attenuation model the ratio of the apparent flux densities should be in one-to-one correspondence with the given ratio of the primary beam model attenuations. From \f{fig:pb} \ it is apparent that this is the case within a few percent on average, thus verifying the primary beam attenuation model used.

\begin{figure}
\includegraphics[width=\columnwidth]{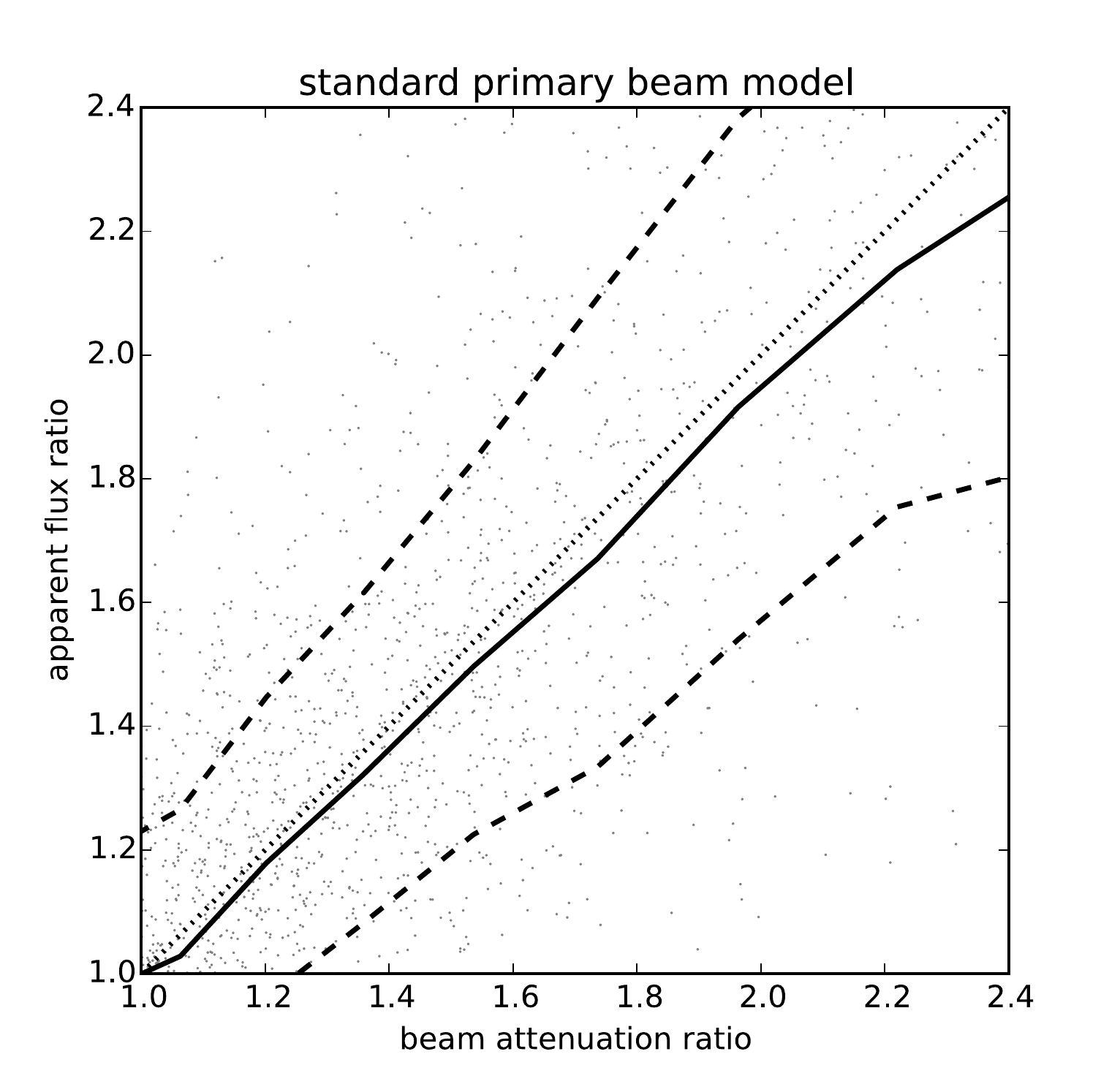}
\protect\protect\caption{
Ratio of  apparent flux density (i.e., not corrected for primary beam attenuation; dots) vs. ratio of the primary beam model attenuations for a sample of bright, compact sources in overlapping pointings. The median and $\pm1\sigma$ offsets are indicated by the full and dashed lines, respectively. The one-to-one line is represented by the dotted line. \label{fig:pb}
}
\end{figure}

%\subsection{Pointing corrections}
%\label{sec:ptcorr}

Following \citet{intema16} we also quantify and apply antenna pointing corrections. In the top panel of \f{fig:ptcorr} \ we show the ratio of flux densities of our  sources in overlapping pointings as a function of the local azimuth of the source position in the first pointing. We account for the deviation from unity (changing sign at about 170 degrees), and in the bottom panel of \f{fig:ptcorr} \ we show the corrected flux density ratios, now consistent with unity.  
We note that the $1\sigma$ scatter of the flux density ratios is $\sim20\%$. As shown in \f{fig:fluxratio} \ this value remains constant as a function of flux density. As the sources used for this analysis have been drawn from overlapping parts of various pointings, i.e. from areas further away from the pointing phase centres, where the $\rm{rms}$ noise is higher and the primary beam corrections less certain (see \f{fig:pb} ), this value should be taken as an upper limit on the relative uncertainty of the source flux densities. 

\begin{figure}
\includegraphics[width=\columnwidth]{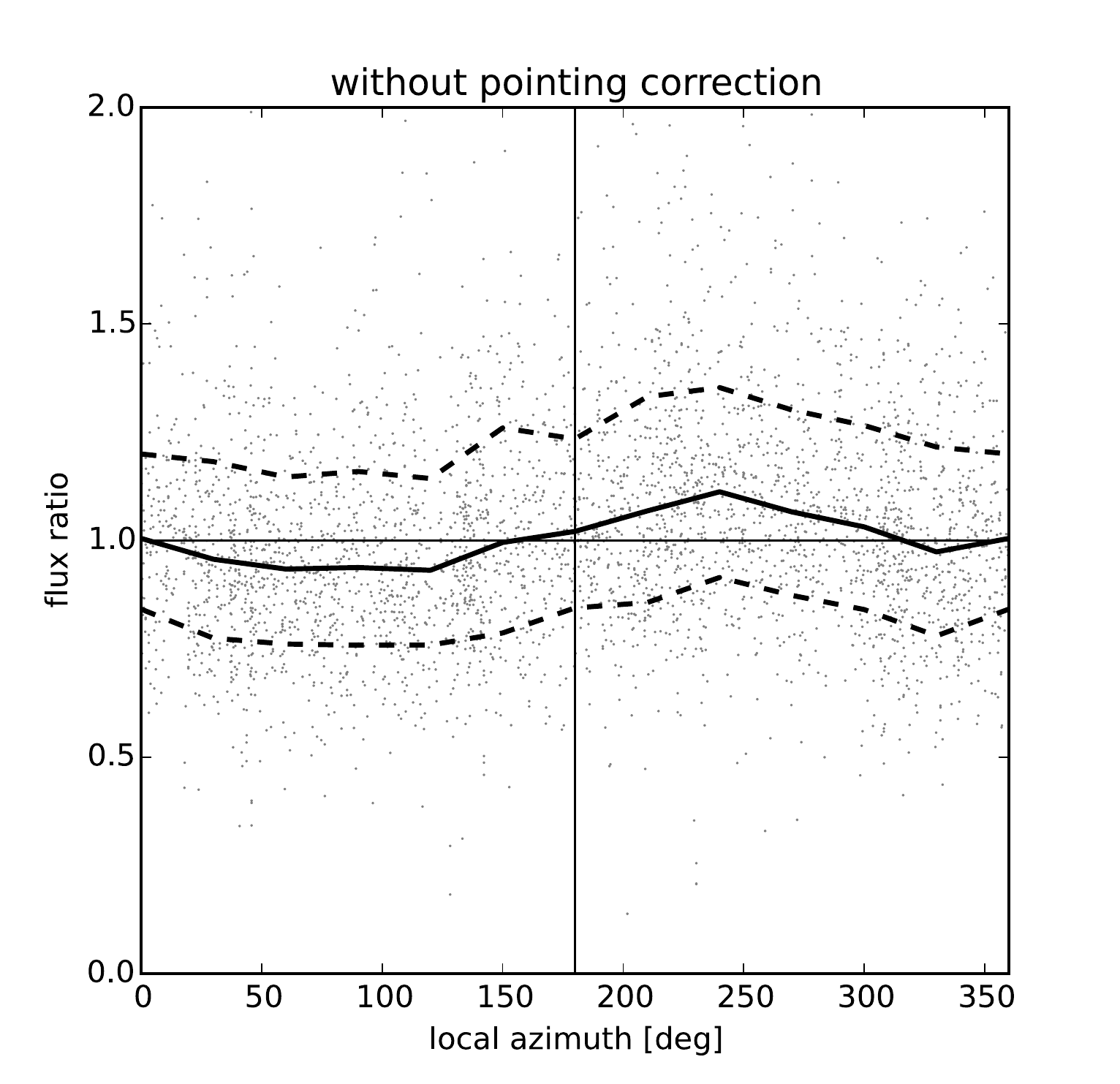}
\includegraphics[width=\columnwidth]{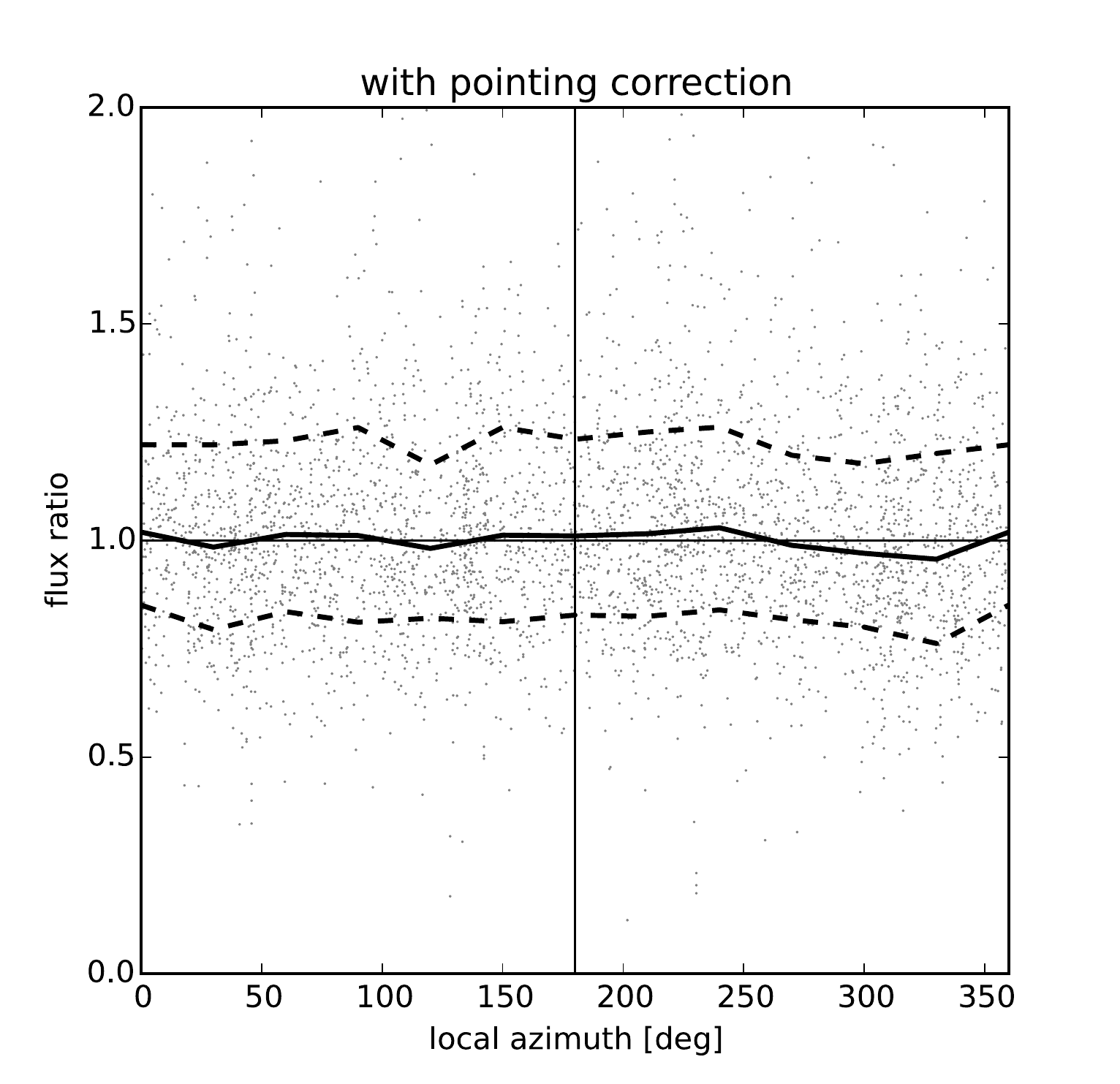}
\protect\protect\caption{
Ratio of flux densities (dots) of bright, compact sources in overlapping pointings as a function of local azimuth of the source position in the first pointing before (top) and after (bottom) applying the pointing correction (see text for details). In both panels the median and $\pm1\sigma$ deviations are indicated by the full, and dashed lines, respectively. \label{fig:ptcorr}
}

\end{figure}
\begin{figure}
\includegraphics[width=\columnwidth]{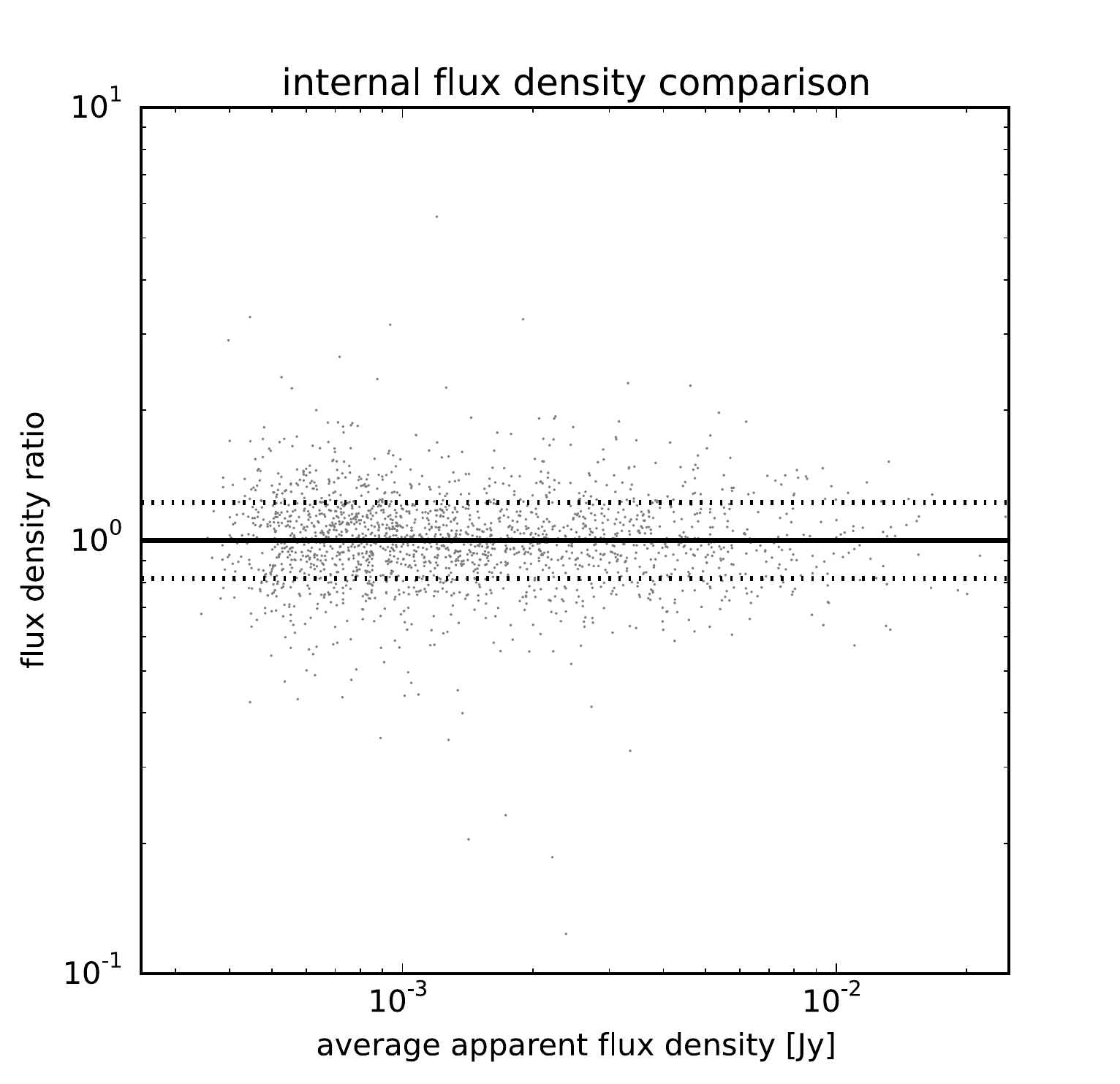}
\protect\protect\caption{
Ratio of flux densities (dots) of bright, compact sources in overlapping pointings as a function of their mean apparent flux density (i.e. uncorrected for primary beam attenuation). The median and $\pm1\sigma$ is indicated by the full and dotted lines, respectively. \label{fig:fluxratio}
}
\end{figure}

\subsection{Final mosaic}
\label{sec:finalmosaic}

After applying the per-pointing astrometric and flux density corrections described above to the individual pointings, each pointing (including its clean component and residual maps) is convolved to a common circular resolution of FWHM $6.5\arcsec\times6.5\arcsec$, prior to mosaicing. This corresponds to a clean beam size larger than or equal to that intrinsically retrieved for the majority of the imaged pointings. The maps are then regridded to $1.9\arcsec\times1.9\arcsec$ pixels, and  combined into a mosaic in such a way that each pixel is weighted as the inverse square of the local rms, estimated using a circular sliding box with a 91-pixel diameter, chosen as a trade-off between minimising false detections at sharp boundaries between high noise and low noise, and separating extended emission regions from high rms regions.
The final mosaic, shown in \f{fig:mosaic} , containing $16,177\times11,493$ pixels, encompasses a total area of $30.4$ square degrees. As can be seen in \f{fig:mosaic} , the rms within the mosaic is highly non-uniform: It decreases from about 200~$\mu$Jy/beam within the XMM-LSS subregion, to about 50~$\mu$Jy/beam in the remaining area. Although  a factor of 3.8 difference in sensitivity is significant, we note that the data processing applied here achieved a background noise reduction of 50\% in the XMM-LSS area relative to the previous data release ($rms\sim300~\mu$Jy/beam; \citealt{tasse07}). For comparison, in  \f{fig:lssxxlmap} \ we show the image of the central part of the XMM-LSS area of the XXL-N field obtained here, and within the previous data release \citep{tasse07}. 

The overall areal coverage of the GMRT-XXL-N mosaic as a function of rms is shown in \f{fig:visibility} . For 60\% of the total 30.4 square degree area an rms better than 150~$\mu$Jy/beam is achieved (with a median value of about 50~$\mu$Jy/beam). For the remainder (corresponding to the XMM-LSS subarea of the XXL-N field) a median rms of about 200~$\mu$Jy/beam is achieved.  
 
\begin{figure*}[ht!]
\centering
\includegraphics[clip, trim=0.cm 7.5cm 0.cm 7.5cm, scale=0.85]{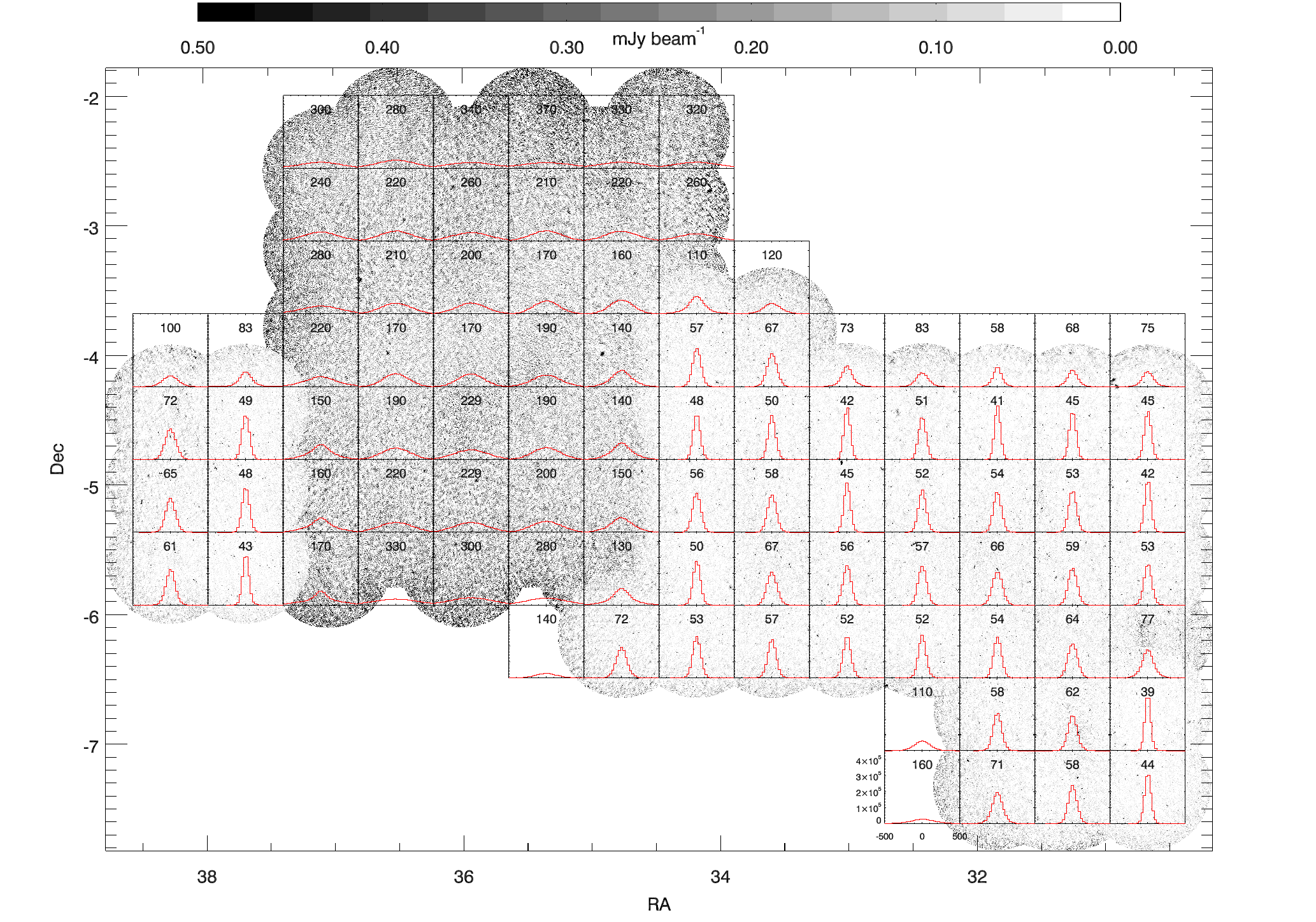}
\protect\protect\caption{
Grayscale mosaic with overlayed pixel flux distributions within small areas, encompassed by the panel (the local rms in $\mu$Jy/beam is indicated in each panel). The x and y ranges for all the panels are indicated  in the bottom left panel. \label{fig:mosaic}
}
\end{figure*}

\begin{figure*}\includegraphics[clip, trim=0.cm 0cm 0.cm 0.cm]{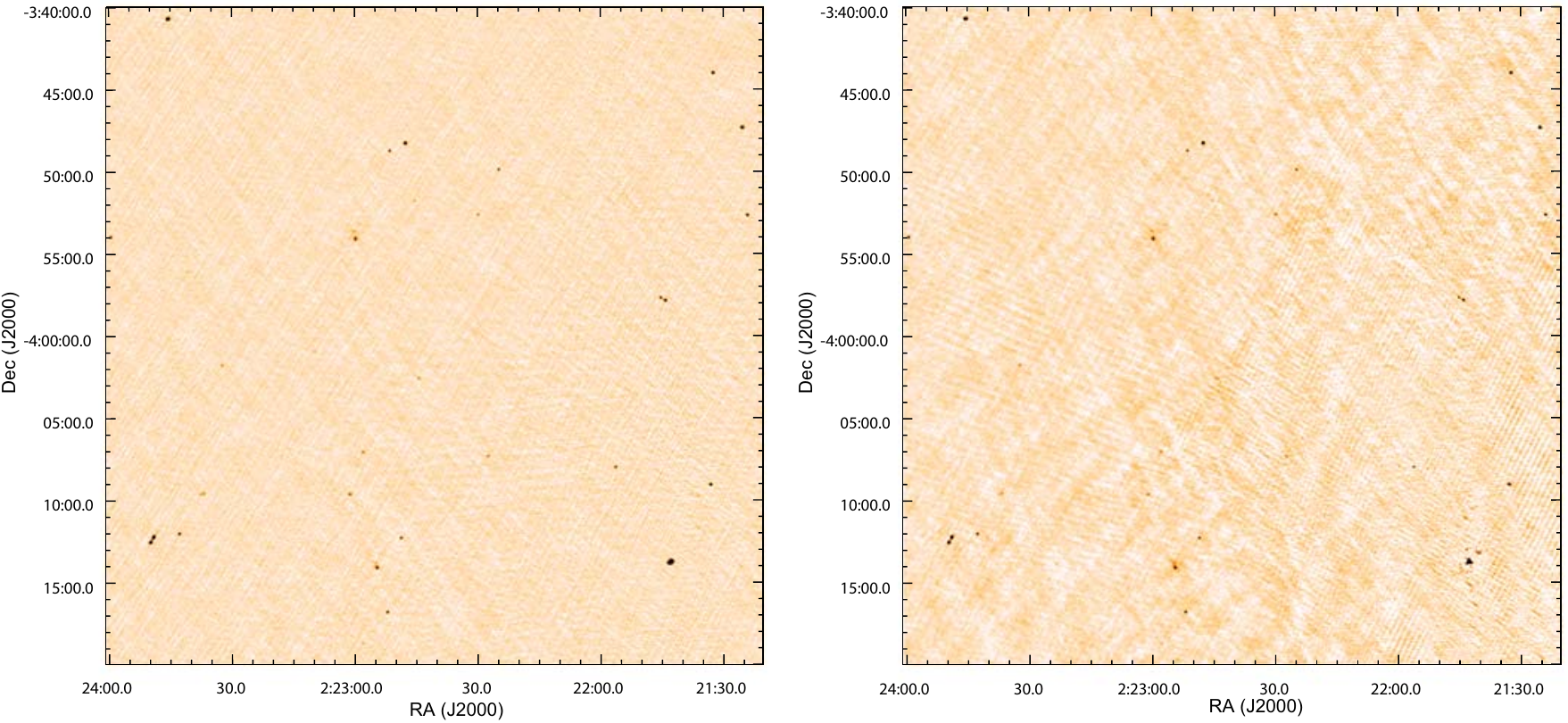}
\protect\protect\caption{
Mosaiced images of the XMM-LSS area of the XXL-N field obtained by the procedure presented here (left panel) and previously published (right panel; \citealt{tasse07}), using the same colour scale.
\label{fig:lssxxlmap}
}
\end{figure*}

\begin{figure}
\includegraphics[clip, trim=0.cm 3cm 0.cm 4.1cm, scale=0.6]{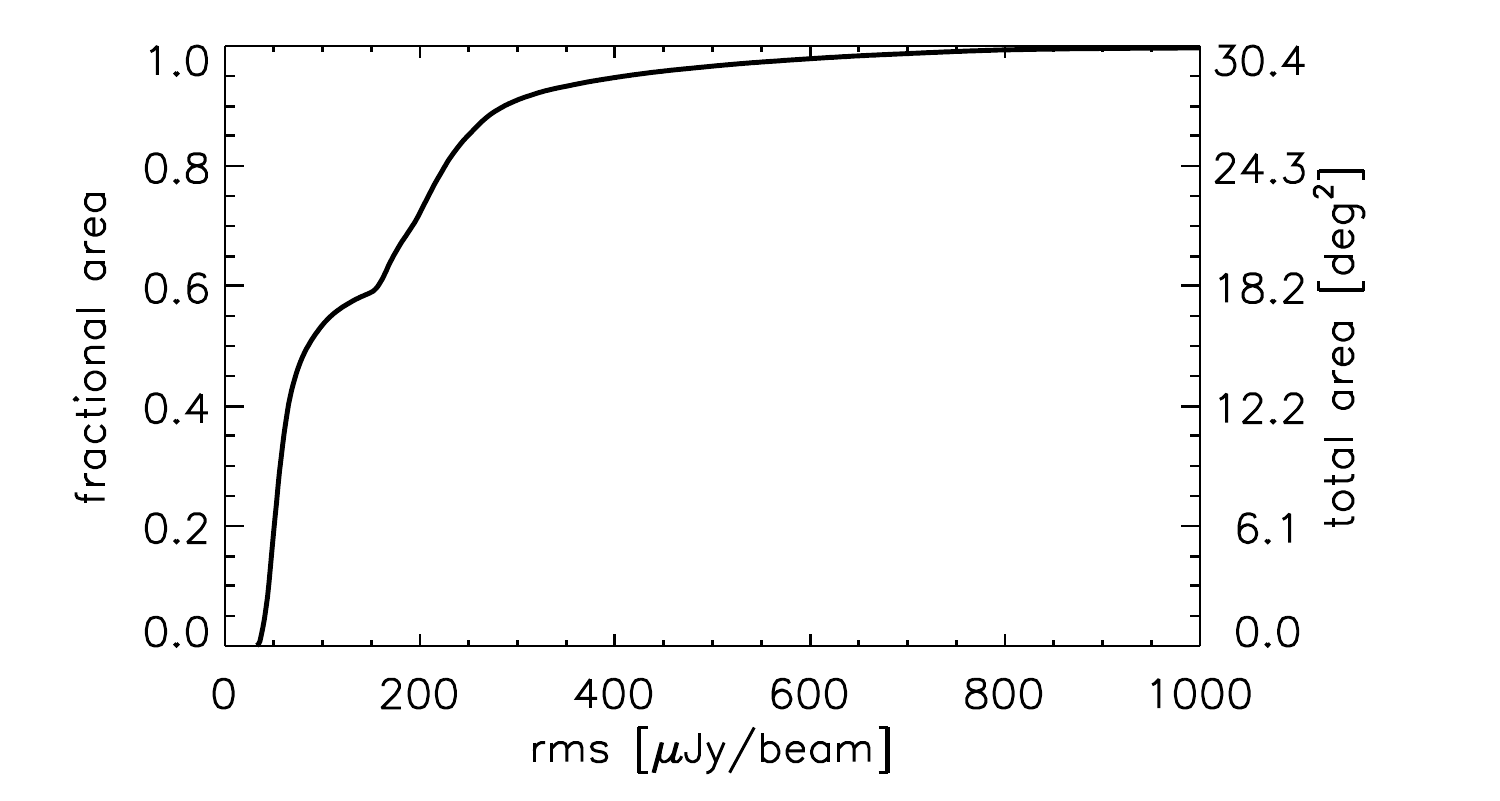}
\protect\protect\caption{
Areal coverage as a function of the rms noise in the mosaic.
\label{fig:visibility}
}
\end{figure}

\section{Cataloguing\label{sec:catalog}}

We describe the source extraction  (\s{sec:pybdsm} ),  corrections performed to account for bandwidth smearing, and the measurement of the  flux densities of resolved and unresolved sources (\s{sec:res} ). We also describe the  process of combining multiple detections,  physically associated with single radio sources (\s{sec:multi} ), and present the final catalogue  (\s{sec:finalcat} ).  

\subsection{Source extraction}
\label{sec:pybdsm}

To extract sources from our mosaic we used the {\sc PyBDSF} software \citep{mohan15}. We set {\sc PyBDSF} to search for islands of pixels with flux density values greater than or equal to three times the local rms noise (i.e $\geq3\sigma$) surrounding peaks above $5\sigma$. %\comm{proper numbers to be put in} . 
To estimate the local rms a box of 195 pixels per side was used, leading to a good trade-off between detecting real objects and limiting false detections (see \s{sec:falsdet} ). Once islands are located, {\sc PyBDSF} fits Gaussian components, and their flux density is estimated after deconvolution of the clean beam. These components are grouped into single sources when necessary, and final source flux densities are reported, as well as flags indicating whether multiple Gaussian fits were performed. The procedure resulted in 5\,434 sources with signal-to-noise ratios $\geq 7$.

\subsection{Resolved and unresolved sources and smearing}
\label{sec:res}

To estimate smearing due to bandwidth- and time-averaging, and to separate resolved from unresolved sources we follow the standard procedure, which relies on a comparison of the sources' total and peak flux densities \citep[e.g.,][]{bondi08,intema14,smo17}. In \f{fig:stotspeak} \ we show the ratio of the total and peak flux densities ($S_\mathrm{T}$ and  $S_\mathrm{P}$, respectively) for the 5\,434  sources as a function of the signal-to-noise ratio ($\mathrm{S/N}\geq7$). While the increasing spread of the points with decreasing S/N ratio reflects the noise properties of the mosaic, smearing effects will be visible as a systematic, positive offset from the $S_\mathrm{T}/S_\mathrm{P}=1$ line as they decrease the peak flux densities, while conserving the total flux densities. To quantify the smearing effect we fit a Gaussian to the logarithm of the $S_\mathrm{T}/S_\mathrm{P}$ distribution obtained by mirroring the lower part of the distribution over its mode (to minimise the impact of truly resolved sources). We infer a mean value of 6\%, and estimate an uncertainty of 1\% based on a range of binning and S/N ratio selections. After correcting the peak flux densities for this effect we fit a lower envelope encompassing 95\% of the sources below the $S_\mathrm{T}/S_\mathrm{P}=1$ line, and mirror it above this line. Lastly, we consider all sources below this curve, of the form 
\begin{equation}
\label{eq:curve}
S_\mathrm{T}/S_\mathrm{P}=1 + 3.2 \times (S/N)^{-0.9},
\end{equation}
as unresolved and set their total flux densities equal to their peak flux densities (corrected for the smearing effects). We note that the four extreme outliers ($S_\mathrm{T}/S_\mathrm{P}<0.5$) are due to blending issues,  a locally high rms, or being possibly spurious.

\begin{figure}
\includegraphics[clip, trim=0.cm 7cm 0.cm 7cm, width=\columnwidth]{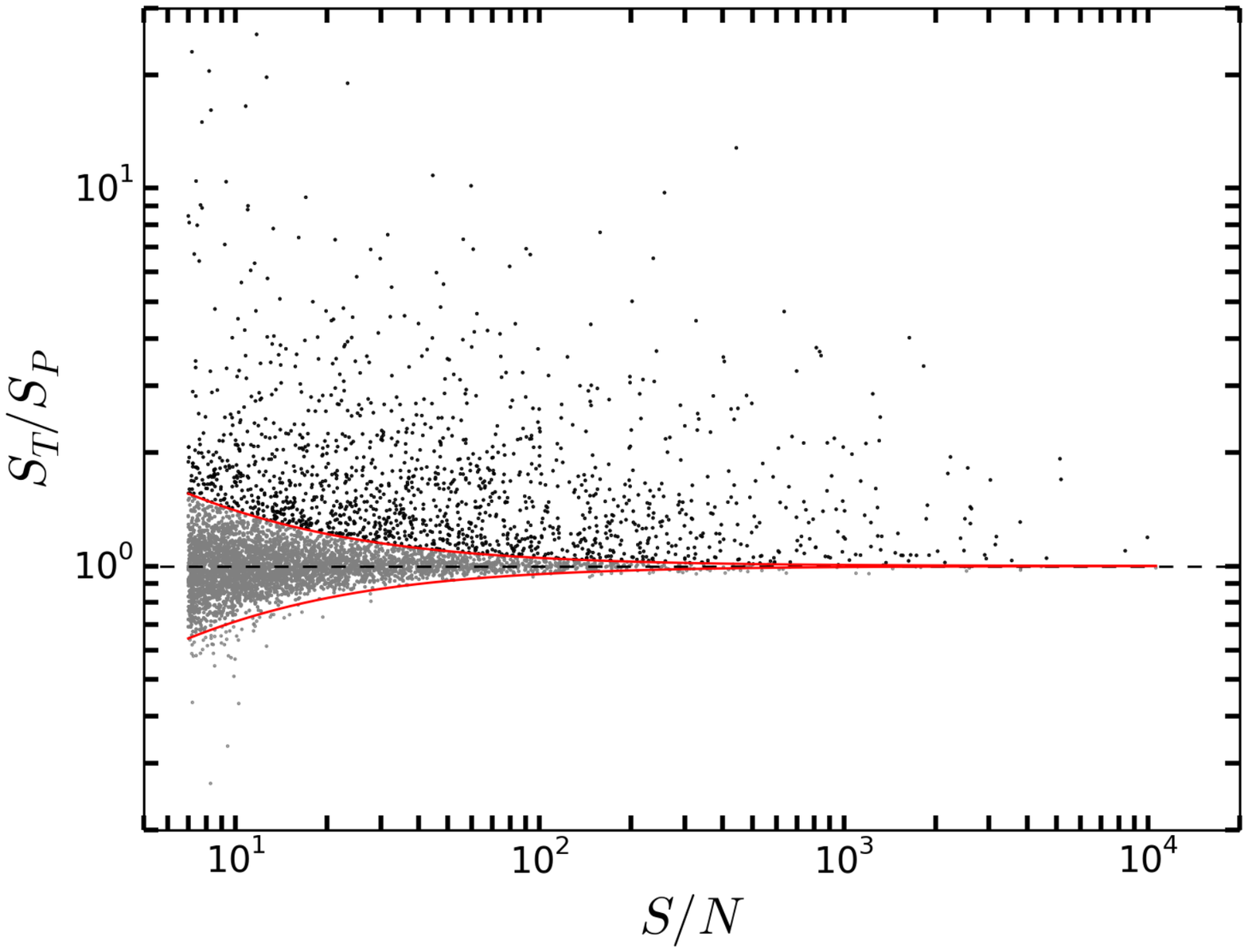}
\protect\protect\caption{
Total ($S_\mathrm{T}$) over peak ($S_\mathrm{P}$) flux as a function of signal-to-noise ratio. The horizontal dashed line shows the $S_\mathrm{T}/S_\mathrm{P}=1$ line. The upper curve was obtained by mirroring the lower curve (which encompasses 95\% of the sources below the $S_\mathrm{T}/S_\mathrm{P}=1$ line; see Eq.~\ref{eq:curve}) over the $S_\mathrm{T}/S_\mathrm{P}=1$ line. All sources below the upper curve are considered unresolved, while those above are considered resolved (see text for details).
\label{fig:stotspeak}
}
\end{figure}

\subsection{Complex sources}
\label{sec:multi}

Radio sources appear in many shapes, and it is possible that  sources with complex radio morphologies (e.g. core, jet, and lobe structures that may be warped or bent) are listed as multiple sources within the source extraction procedure. To identify these sources we adopt the procedure outlined in \citet{tasse07}. We identify groups of components within a radius of $60\arcsec$ from each other. Based on the source density in the inner part of the field, the Poisson probability is 0.22 $\%$ that two components are associated by chance. For the outer part of the field, an additional flux limit of $S_\mathrm{610MHz}>1.4$~mJy is imposed on the components prior to identifying  the groups they belong to. This flux limit is justified by the size-flux relation for radio sources \citep[e.g.][]{bondi03}, i.e. brighter sources are larger in size and are thus more likely to break into multiple components, and it also assures a Poisson probability of 0.39 $\%$ for a chance association. All of these groups were visually checked a posteriori, and verified against an independent visual classification of multicomponent sources performed by six independent viewers. In total we identify 768 sources belonging to  337 distinct groups of multiple radio detections likely belonging to a single radio source (see also next section).

%a visual inspection of all sources against the optical/NIR CFHTLS images was performed by six independent viewers. We have identified xx sources with complex radio morphologies, listed as multiple entries in the {\sc PyBDSF} extracted catalog. We have further corrected the catalog entries such that the multiple entries were combined into single entries. 

\subsection{Final catalogue}
\label{sec:finalcat}

In our final catalog  for each source we report the following:
\begin{itemize}
\item[ ]{Column 1:} Source ID;\vspace{2mm}

\item[ ]{Columns 2-5:} RA and Dec position (J2000) and error on the position as provided by {\sc PyBDSF};\vspace{2mm}

\item[ ]{Column 6:} Peak flux density ($S_\mathrm{P}$) in units of Jy/beam, corrected for smearing effects as detailed in \s{sec:res} ;\vspace{2mm}

\item[ ]{Column 7:} Local rms value in units of Jy/beam; \vspace{2mm}

\item[ ]{Column 8:} S/N ratio;\vspace{2mm}

\item[ ]{Column 9 - 10:} Total flux density ($S_\mathrm{T}$) and its error, in units of Jy;\vspace{2mm}

\item[ ]{Column 11:} Flag for resolved sources; 1 if resolved, 0 otherwise. We note that the total flux density for unresolved sources equals the smearing corrected peak flux and the corresponding error lists rms scaled by the correction factor (see \s{sec:res} \ for details); \vspace{2mm}

\item[ ]{Column 12:} Complex source identifier, i.e., ID of the group obtained by automatic classification of 
                     multicomponent sources (see \s{sec:multi} \ for details); if 0 no group associated with the source, otherwise the integer corresponds to the group ID;\vspace{2mm}
                    
\item[ ]{Column 13:} Spectral index derived using the 610 MHz and 1.4 GHz (NVSS) flux densities, where available (-99.99 otherwise); \vspace{2mm}
                     
\item[ ]{Column 14:} Area flag; $0$ if the source is in the inner mosaic area (within the XMM-LSS field and with higher rms), $1$ if it is in the outer field area (with better rms; see \s{sec:finalmosaic} \ and \f{fig:mosaic} \ for details);\vspace{2mm}

\item[ ]{Column 15:}  Edge flag; 0 if the source is on the edge where the noise is high, $1$ otherwise. We note that selecting this flag to be 1 in the inner (outer) area, i.e. for area flag 0 (1) selects sources within areas of 7.7 (12.66) square degrees, while the area corresponds to 9.63 square degrees for the inner area (area flag 0) and an edge flag of 0 or 1. \vspace{2mm}
\end{itemize}
%its ID, the RA and DEC position and error on the position, the peak flux density corrected for smearing effects, the local rms value, S/N ratio, total flux density with its relative error, the flag for resolved sources, the flag for multi-component sources, and the flag for the source's presence in the XMM-LSS or the outer XXL-N area. We note that the total flux densities for unresolved sources equal their (smearing corrected) peak flux densities. 
The full catalogue is available as a queryable database table
(XXL$\_$GMRT$\_$17) via the XXL Master Catalogue browser\footnote{\url{http://cosmosdb.iasf-milano.inaf.it/XXL}}. A copy will also be deposited at the
Centre de Donn\'ees astronomiques de Strasbourg (CDS)\footnote{\url{http://cdsweb.ustrasbg.fr}}.

\section{Reliability and completeness}
\label{sec:tests}

In this section we assess the astrometric accuracy and false detection rate within the catalogue presented above (\s{sec:mosaicastrom} \ and \s{sec:falsdet} , respectively). We compare the source flux densities to those derived from the previous XMM-LSS GMRT 610 MHz data release (\s{sub:Flux-comparison} ). We then  derive average 610 MHz -- 1.4 GHz spectral indices for unbiased subsamples of bright sources (\s{sec:alpha} ), and use these to construct  1.4~GHz radio source counts, which we compare to counts from much deeper radio continuum surveys and thereby assess and quantify the incompleteness of our survey as a function of flux density (\s{sec:counts} ).

\subsection{Astrometric accuracy}
\label{sec:mosaicastrom}

To assess the astrometric accuracy of the presented mosaic and catalogue, we compared the positions of our compact sources with those given in the FIRST survey. Using a search radius of $5\arcsec$ we found 1\,286 matches. The overall median offset is (as expected) small ($\Delta\mathrm{RA}=-0.01\arcsec$, $\Delta\mathrm{Dec}=0.01\arcsec$), and the 1$\sigma$ position scatter radius is $0.62\arcsec$. This is similar to the internal positional accuracy seen in pointing overlap regions, and corresponds to roughly one-tenth of the angular resolution in the mosaic ($6.5\arcsec$).

\begin{figure}
\includegraphics[width=\columnwidth]{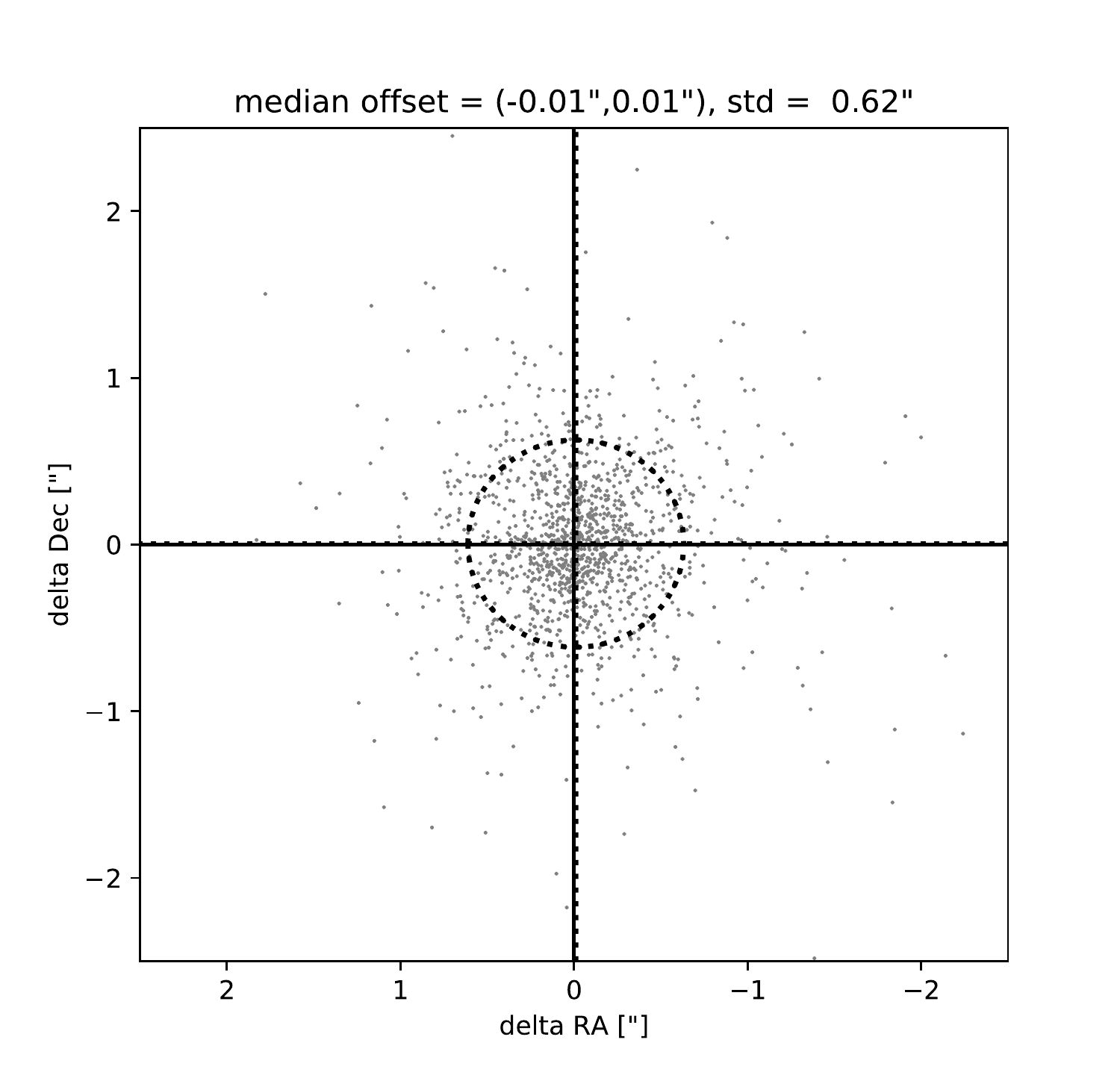}
\protect\protect\caption{
Positional offsets of sources detected in the XXL mosaic, relative to those catalogued within the FIRST survey. The median offsets in RA and Dec are indicated by the vertical and horizontal lines, respectively, while the dotted circle represents the $1\sigma$ deviation (also indicated at the top).
\label{fig:astromfinal}
}
\end{figure}

\subsection{False detection rate}
\label{sec:falsdet}

To assess the false detection rate, i.e. estimate the number of spurious sources in our catalogue, we ran {\sc PyBDSF} on the inverted (i.e. multiplied by $-1$) mosaic. Each 'detection' in the inverted mosaic can be considered spurious as no real sources exist in the negative part of the mosaic. Running {\sc PyBDSF} with the same set-up as for the catalogue presented in \s{sec:catalog} \ , we found only one detection with S/N ratio $\geq7$ in the inverted mosaic. Since there are 5\,434 detections with $\mathrm{S/N}\geq7$  in the real mosaic, false detections are not significant ($<10^{-3}$ of the source population).

\subsection{Flux comparison within the XMM-LSS area}
\label{sub:Flux-comparison}

\begin{figure}
\includegraphics[clip, trim=0.cm 7cm 0.cm 7cm, width=\columnwidth]{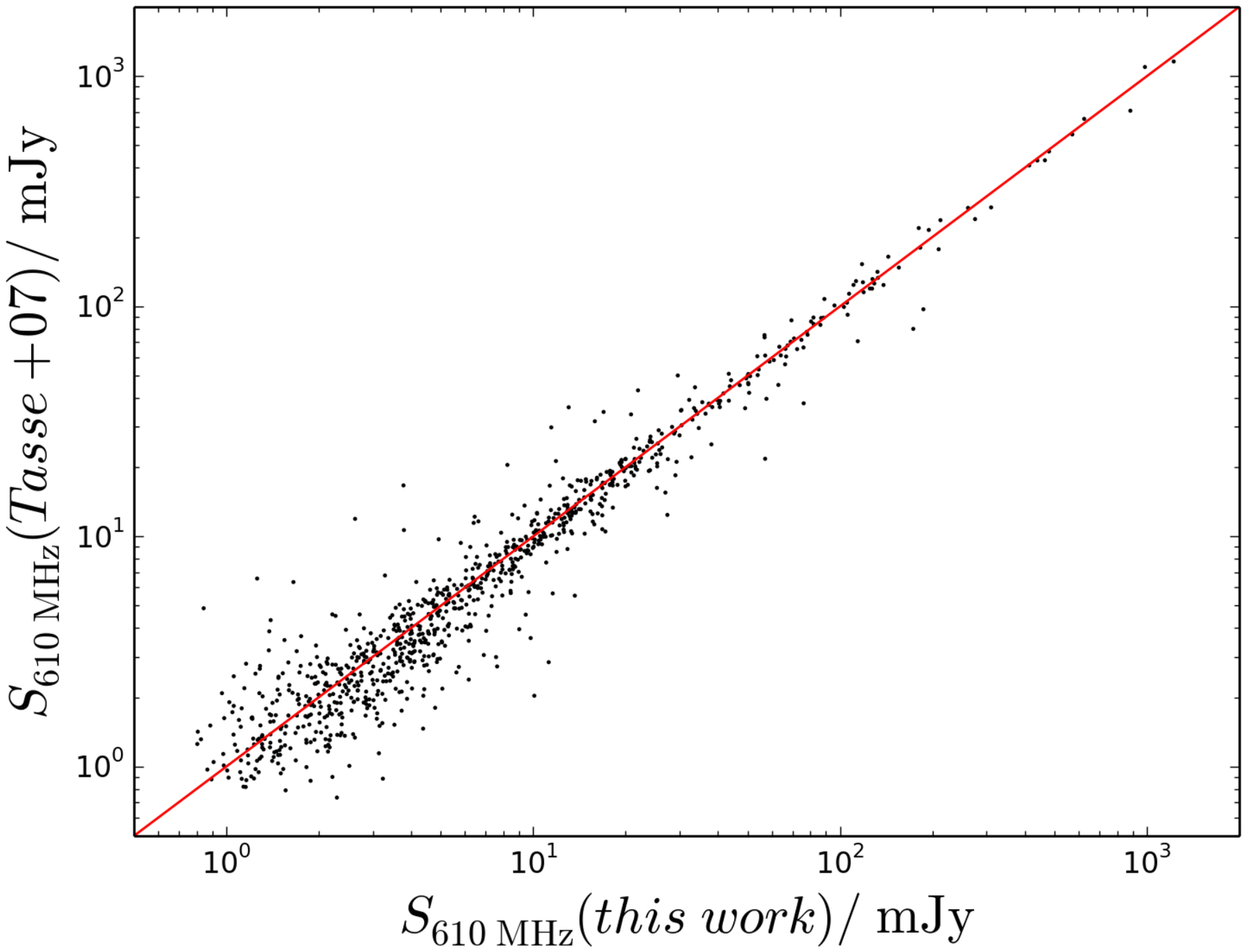}
\includegraphics[clip, trim=0.cm 7cm 0.cm 7cm, width=\columnwidth]{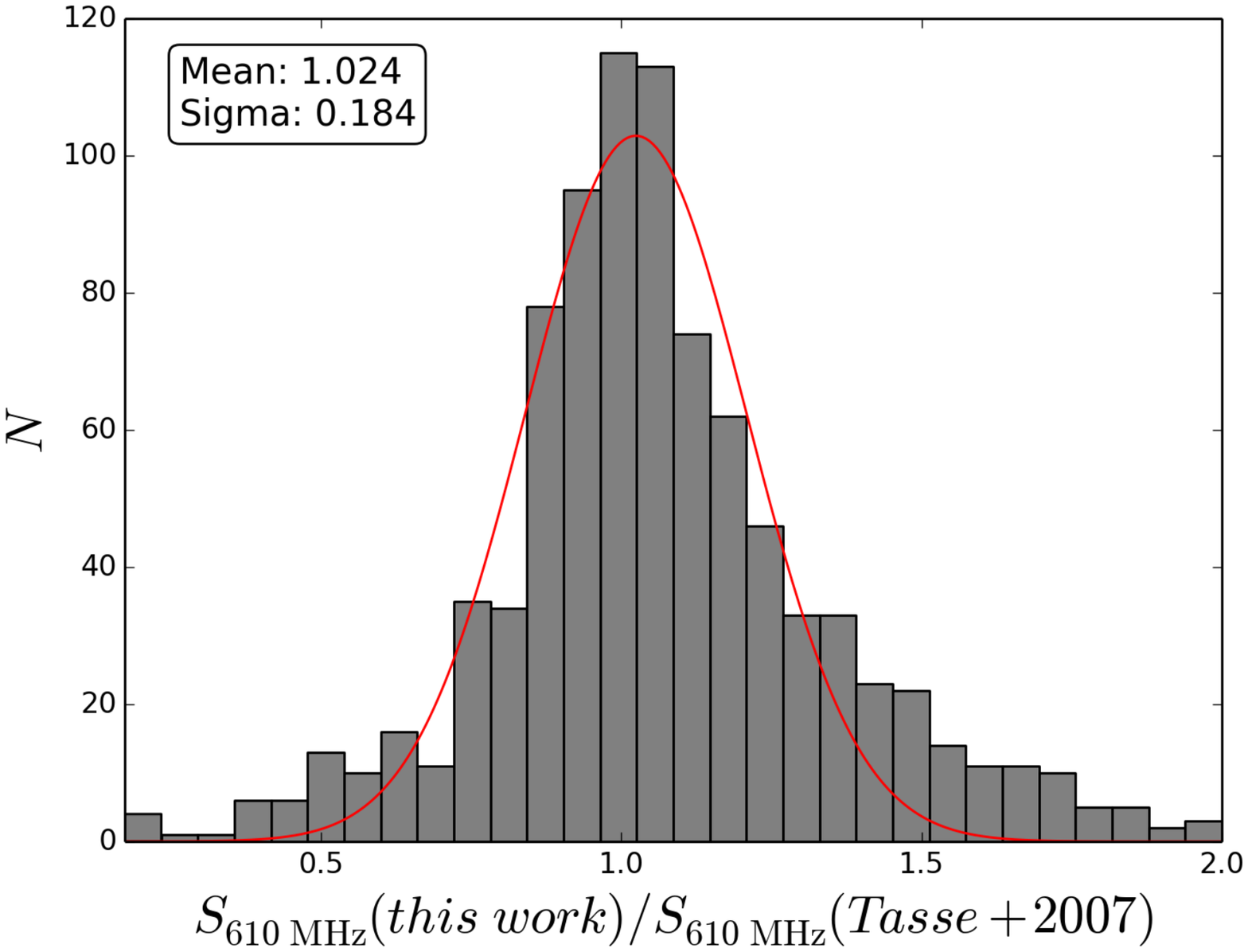}
\protect\protect\caption{
Comparison within the XMM-LSS area of the XXL-N field between the flux densities  obtained here (x-axis) and those extracted in the same way, but over the mosaic published by \citet[][y-axis; top panel]{tasse07}. The solid line is the diagonal. The distribution of the flux ratio, with fitted Gaussian is shown in the bottom panel. 
\label{fig:lssxxlflux}
}
\end{figure}

We compared the flux densities for our sources extracted within the XMM-LSS area of the XXL-N field with those extracted in the same way, but over the XMM-LSS mosaic published by \citet{tasse07}. Using a search radius of $5\arcsec$ we find 924 sources common to the two catalogues. In Fig.~\ref{fig:lssxxlflux} we compare the {\sc PyBDSF}-derived flux densities (prior to corrections for bandwidth smearing, and total flux densities). Overall, we find good agreement, with an average offset of 2.4\%. This can be due to the use of different flux standards (\citealt{perley13} and \citealt{scaife12}). We note that the comparison remains unchanged if flux densities from the \citet{tasse07} catalogue are used instead. %\comm{do we want to say something about the scatter?}

\subsection{Spectral indices}
\label{sec:alpha}

\begin{figure}
\includegraphics[clip, trim=0.cm 0.cm 0.cm 1.5cm, width=1\columnwidth]{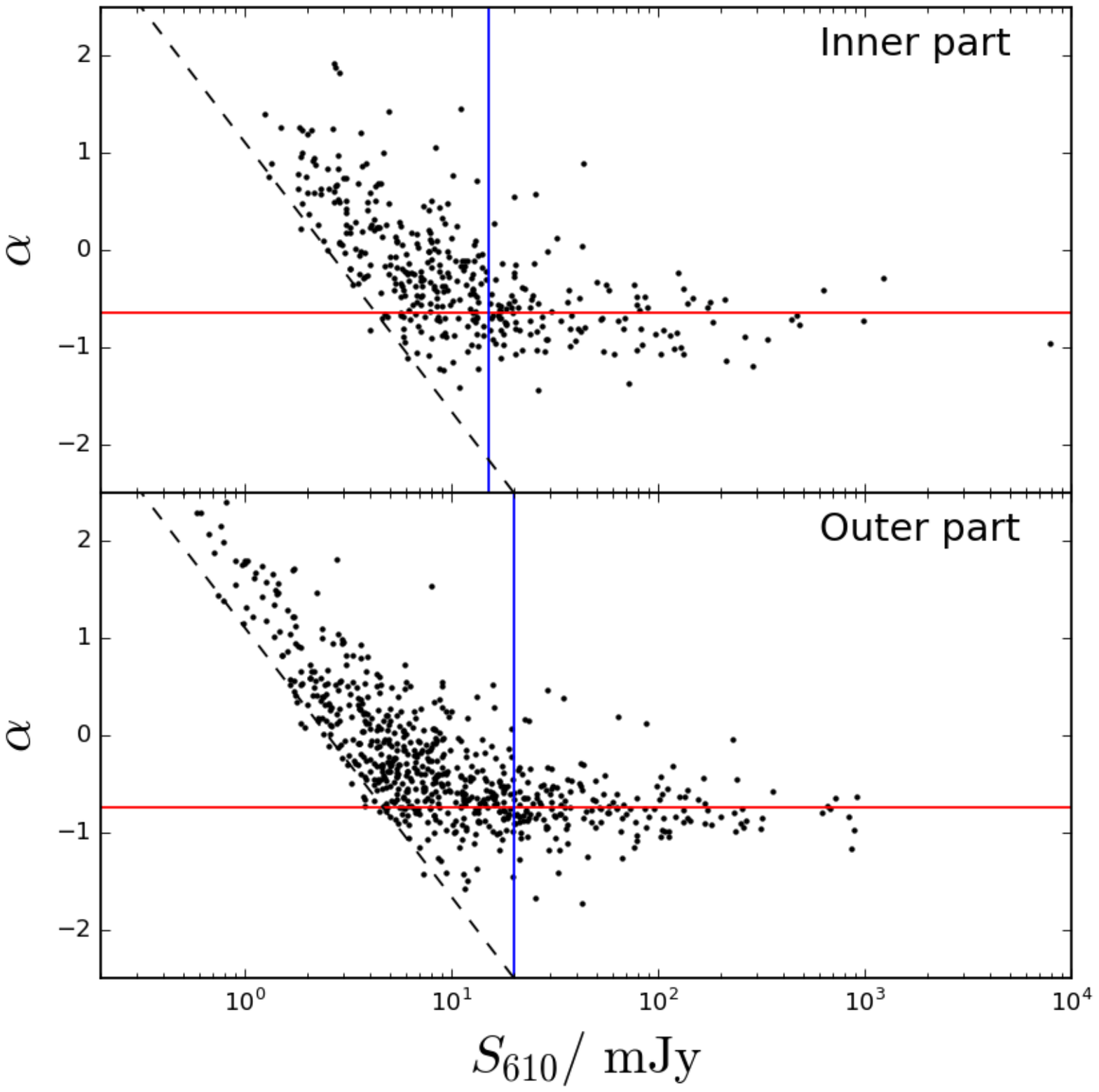} 
\protect\protect\caption{
Spectral index based on 610 MHz and 1.4 GHz (NVSS) data ($\alpha$) as a function of 610~MHz flux density, separately shown for the XMM-LSS (top panel), and outer XXL-N (bottom panel) areas. The dashed line in both panels indicates the constraint on $\alpha$ placed by the NVSS detection limit (2.5~mJy), and the vertical full line indicates the threshold beyond which the sample is not expected to be biased by the different detection limits. The horizontal line indicates the average spectral index for sources with flux densities above this threshold.
\label{fig:alpha}
}
\end{figure}

We derived the 610~MHz -- 1.4 GHz spectral indices for the  sources in our catalogue that are also detected in the the NVSS survey \citep{condon98}. We found 1\,395 associations (470 within the XMM-LSS subarea) using a search radius of $20\arcsec$, which corresponds to about half of the NVSS synthesised beam (FWHM~$\sim45\arcsec$), and assures minimal false matches. In Fig.~\ref{fig:alpha} we show the derived spectral indices as a function of 610~MHz flux density, separately for the XMM-LSS and the outer XXL areas (given the very different rms reached in the two areas). The different source detection limits of the NVSS and GMRT-XXL-N surveys would bias the derived spectral indices if taken at face value. Thus, to construct unbiased samples we defined flux density cuts of  
$S_\mathrm{610MHz}\geq20$~mJy and $\geq2$~mJy for the XMM-LSS, and outer XXL areas, respectively. As illustrated in Fig.~\ref{fig:alpha}, these cuts conservatively assure samples with unbiased $\alpha$ values. For samples defined in this way we find average spectral indices of -0.65 and -0.75 with standard deviations of 0.36 and 0.34, respectively. The values correspond to typically observed spectral indices of radio sources at these flux levels \citep[e.g.,][]{condon92, kimball08}, and they are consistent with those inferred by \citet{tasse07} based on the previous (XMM-LSS) data release.

\subsection{Source counts and survey incompleteness}
\label{sec:counts}

In \f{fig:counts} \ we show the Euclidean normalised source counts at 1.4~GHz frequency derived for our GMRT-XXL-N survey, separately for the inner (XMM-LSS) and outer XXL-N areas. We have chosen a reference frequency of 1.4~GHz for easier comparison with the counts from the literature, which are deeper than our data and are based on data and on simulations \citep{condon84,wilman08, smo17}. The 610 MHz flux densities were converted to 1.4~GHz using a spectral index of $\alpha=-0.7$, as derived in the previous section. The area considered for deriving the GMRT-XXL-N counts within the inner (XMM-LSS) area (with an rms of $\sim200~\mu$Jy/beam) was 9.6 square degrees (green symbols in \f{fig:counts} ). Two areas were considered for the outer XXL-N region: the full area of 20.75 square degrees (also containing the noisy edges; see \f{fig:mosaic} ) and an area of 12.66 square degrees, which excludes the noisy edges and is characterised by a fairly uniform rms of $\sim45~\mu$Jy/beam (black and yellow symbols, respectively, in \f{fig:counts} ). We find that the counts within these 12.66 square degrees  are $40-60$\% higher than those derived using the full outer area (below 0.4 mJy). This suggests that including the noisy edges (as would be expected) further contributes to survey incompleteness; hereafter we thus only consider the 12.66 square degree area for the outer GMRT-XXL-N field.

In \f{fig:counts} \ we also show the 1.4~GHz counts derived by \citet{condon84}, \citet{wilman08} and \citet{smo17}. \citet{condon84} developed a  model for the source counts using the 
local 1.4 GHz luminosity function for two dominant, spiral and elliptical galaxy populations combined with source counts, redshift, and spectral-index distributions for various 400 MHz to 5 GHz
flux limited samples. The Square Kilometre Array Design Study (SKADS) simulations of the radio source counts were based on evolved luminosity functions of various radio populations, also accounting for large-scale clustering \citep{wilman08}. Lastly, the counts taken from \citet{smo17} were constructed using the VLA-COSMOS 3~GHz Large Project, to date the deepest ($\mathrm{rms}\sim2.3~\mu$Jy/beam) radio survey of a relatively large  field (2 square degrees). Their 1.4 GHz counts were derived using the average spectral index ($\alpha=-0.7$) inferred for their 3 GHz  sources. We note that the various counts from the literature are consistent down to the flux level reached by our GMRT-XXL-N data ($\sim0.15~$mJy), and that they are in good agreement with our GMRT counts down to $\sim2$~mJy, implying that our survey is complete at 1.4 GHz flux densities $>2$~mJy. This corresponds to 610 MHz  flux densities of  $>4$~mJy, beyond which we take the detection fractions and completeness to equal unity (see below).

The decline at 1.4 GHz (610 MHz) flux densitites $\lesssim2$~mJy ($\lesssim4$~mJy) of the derived GMRT-XXL-N counts, compared to the counts from the literature (see \f{fig:counts} ) , can be attributed to the survey incompleteness. In radio continuum surveys this is due to a combination of effects, such as real sources remaining undetected due to their peak brightnesses falling below the detection threshold given the noise variations across the field, or source extendedness, and the flux densities of those  detected being over- or underestimated due to these noise variations. Commonly, such survey incompleteness is accounted for by statistical corrections (as a function of flux density) taking all these combined effects into account \citep[e.g.][]{bondi08,smo17}. The approach we take here makes use of  
 the availability of radio continuum surveys that reach much deeper than our GMRT-XXL-N 610 MHz survey. In particular, for the inner (XMM-LSS; 9.6 square degrees, $\mathrm{rms}\sim200~\mu$Jy/beam) and outer (12.66 square degrees, $\mathrm{rms}\sim45~\mu$Jy/beam) GMRT-XXL-N areas, we derive the measurements relative to the VLA-COSMOS 3 GHz source counts. The $\sim2.3~\mu$Jy/beam depth of the VLA-COSMOS 3 GHz Large Project means these counts are 100\% complete in the flux regime encompassed by the GMRT-XXL-N survey (see  \citealt{smo17} for details). 
 
 The GMRT-XXL-N survey differential incompleteness measurements (i.e. detection fractions as a function of flux density) are  computed as the ratio of the GMRT-XXL-N and VLA-COSMOS source counts, are shown in \f{fig:compl} , and listed in \t{tab:compl} \ separately for the  inner and outer areas of the mosaic. 
  The listed corrections, combined with the 'edge flag', given in the catalogue (see \s{sec:finalcat} ), can be used to statistically correct for the incompleteness of the source counts important to determine luminosity functions. 
 In \f{fig:compl} \ we also show the total completeness of the survey within the given areas corresponding to the fraction of sources detected above the given flux bin (lower) limit; in \f{fig:compl2} \ we show the same, but for the full 30.4 square degree mosaic. We note that we consider the survey complete beyond $S_\mathrm{610MHz}=4.6$~mJy, below which the detection fraction decreases. The overall completeness for the full area (30.4 square degree survey) reaches 50\% at $S_\mathrm{610MHz}\approx400~\mu$Jy (i.e. about $250~\mu$Jy in the outer XXL area and about $900~\mu$Jy in the inner XXL area). Furthermore, comparing the detection fraction derivation with that derived relative to the SKADS simulation, we estimate a possible uncertainty of the derived values of the order of 10\%.

\begin{figure*}
\begin{center}
\includegraphics[clip, trim=0.8cm 9cm 0.cm 8cm, width=1.56\columnwidth]{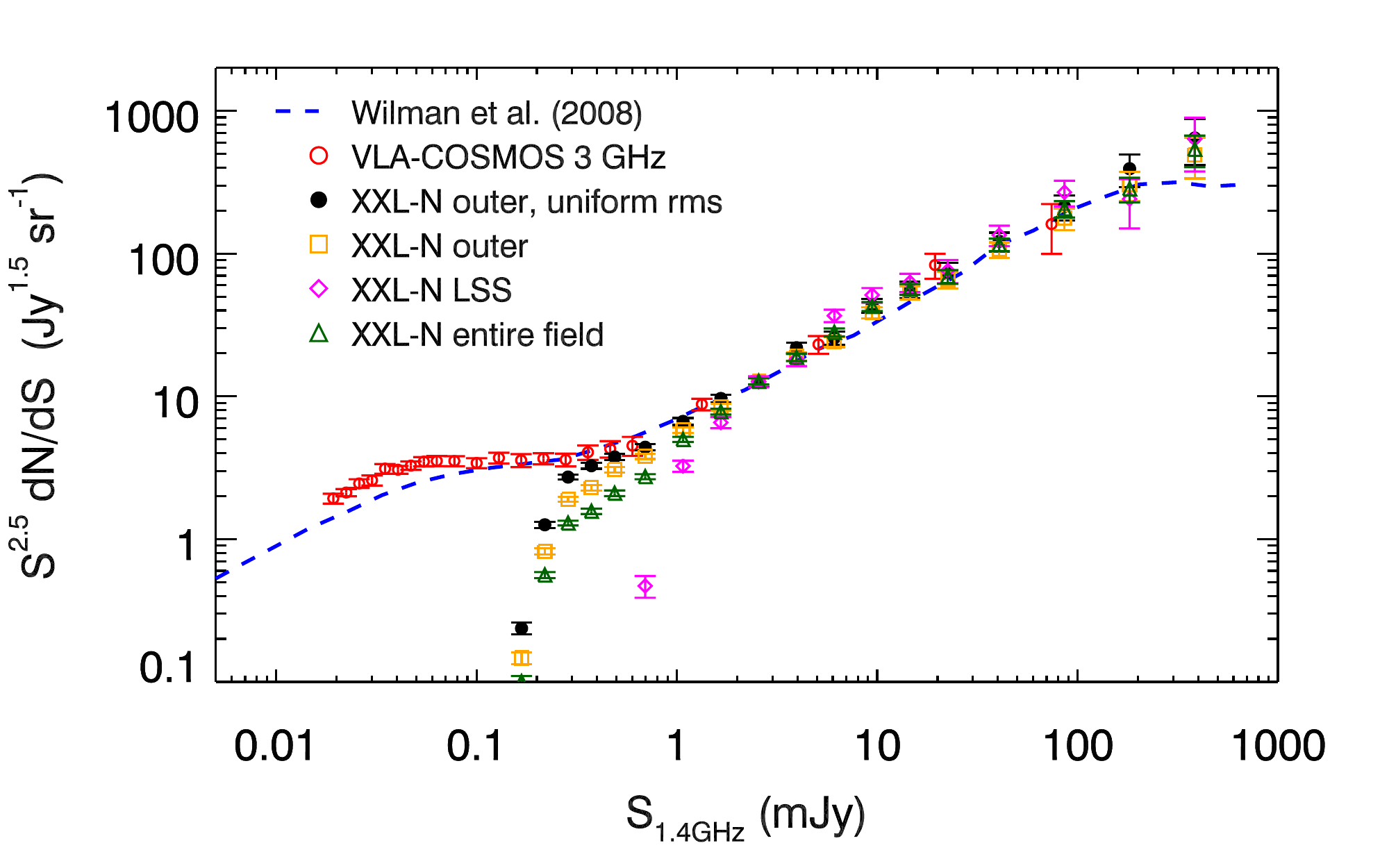}
\protect\protect\caption{
Source counts at 1.4~GHz , normalised to Euclidean space separately for various surveys (symbols) and simulations (lines), as indicated in the panel. The 610 MHz flux densities have been translated to 1.4 GHz using a spectral index of $\alpha=-0.7$, as derived in \s{sec:alpha} . The rapid decline of the GMRT-XXL-N counts (at $S_\mathrm{1.4GHz}<2$~mJy) is due to survey incompleteness (see text for details). 
\label{fig:counts}
}
\end{center}
\end{figure*}

\begin{figure}
\includegraphics[clip, trim=0.cm 8.5cm 0.cm 9cm, width=1.06\columnwidth]{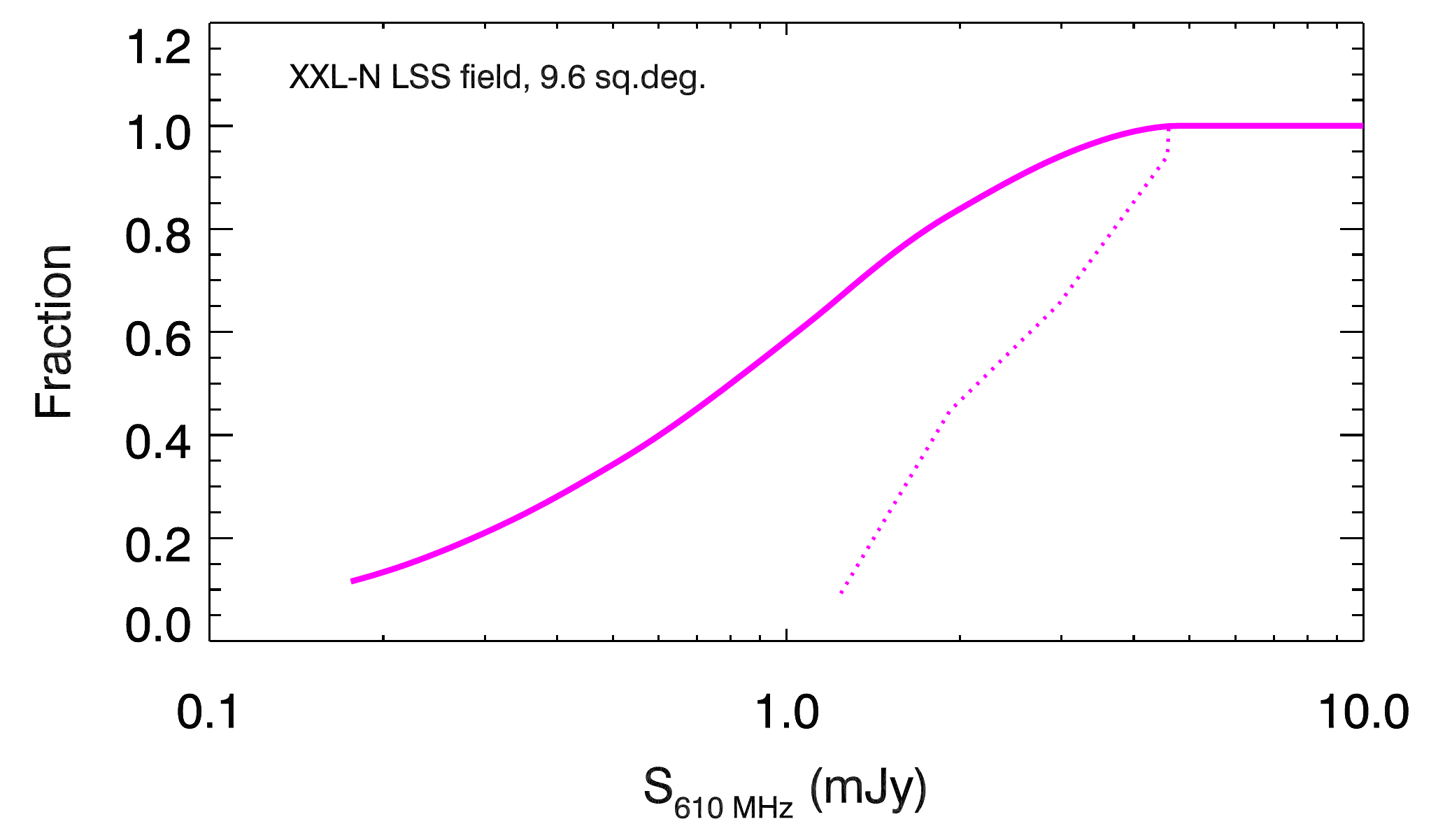}
\includegraphics[clip, trim=0.cm 8.5cm 0.cm 9cm, width=1.06\columnwidth]{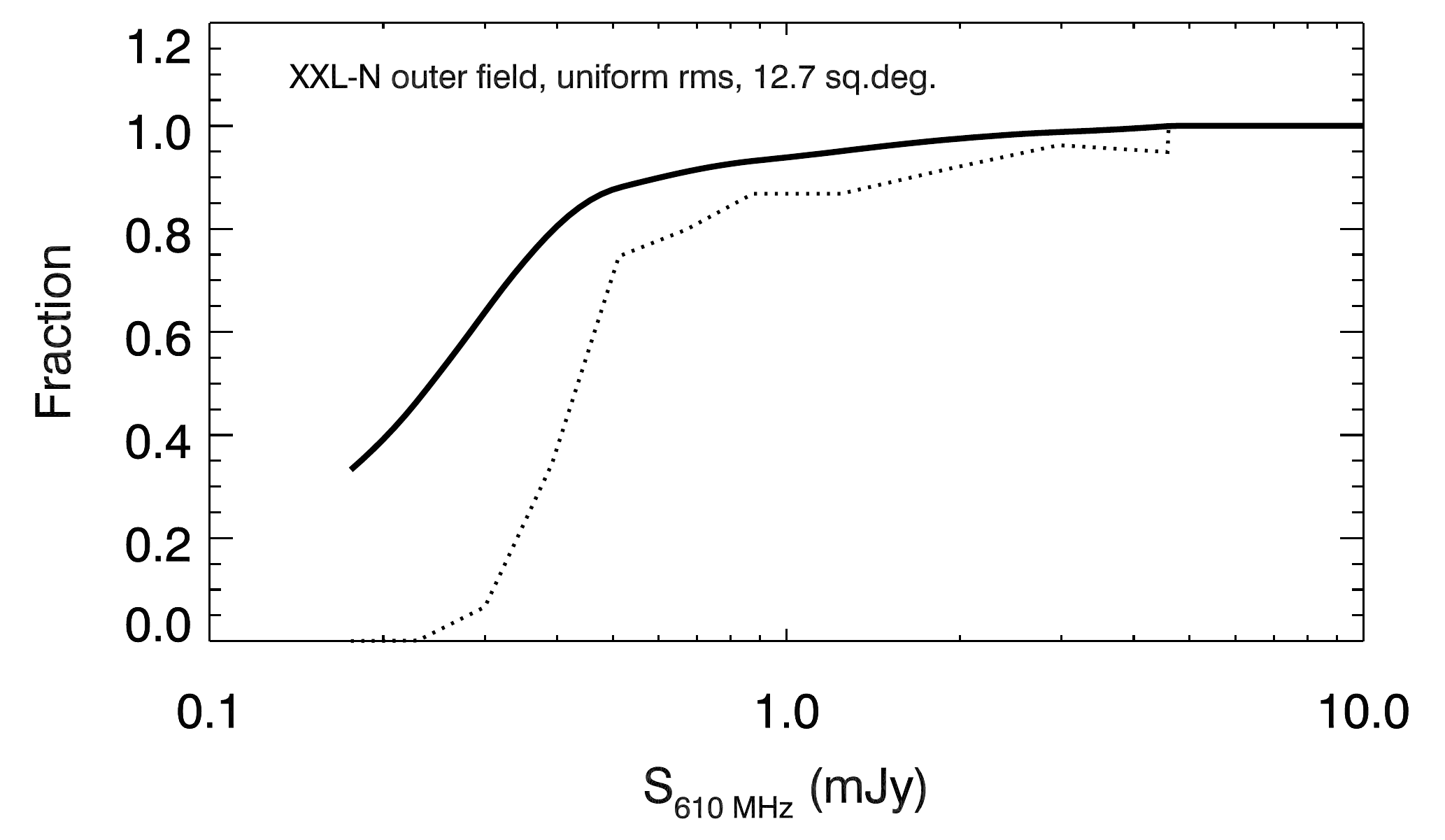}
\protect\protect\caption{
Detection fraction as a function of 610~MHz flux density relative to the VLA-COSMOS 3~GHz Large Project (dashed line) for the inner area (9.6 square degrees, coincident with the XMM-LSS field; left panel) and outer area (12.66 square degree) areas. Also shown is the completeness (i.e. the detection fraction in a given bin and that above the given flux bin lower limit) as a function of flux density (full line). 
\label{fig:compl}
}
\end{figure}

\begin{figure}
\includegraphics[clip, trim=0.cm 8.5cm 0.cm 9cm, width=1.06\columnwidth]{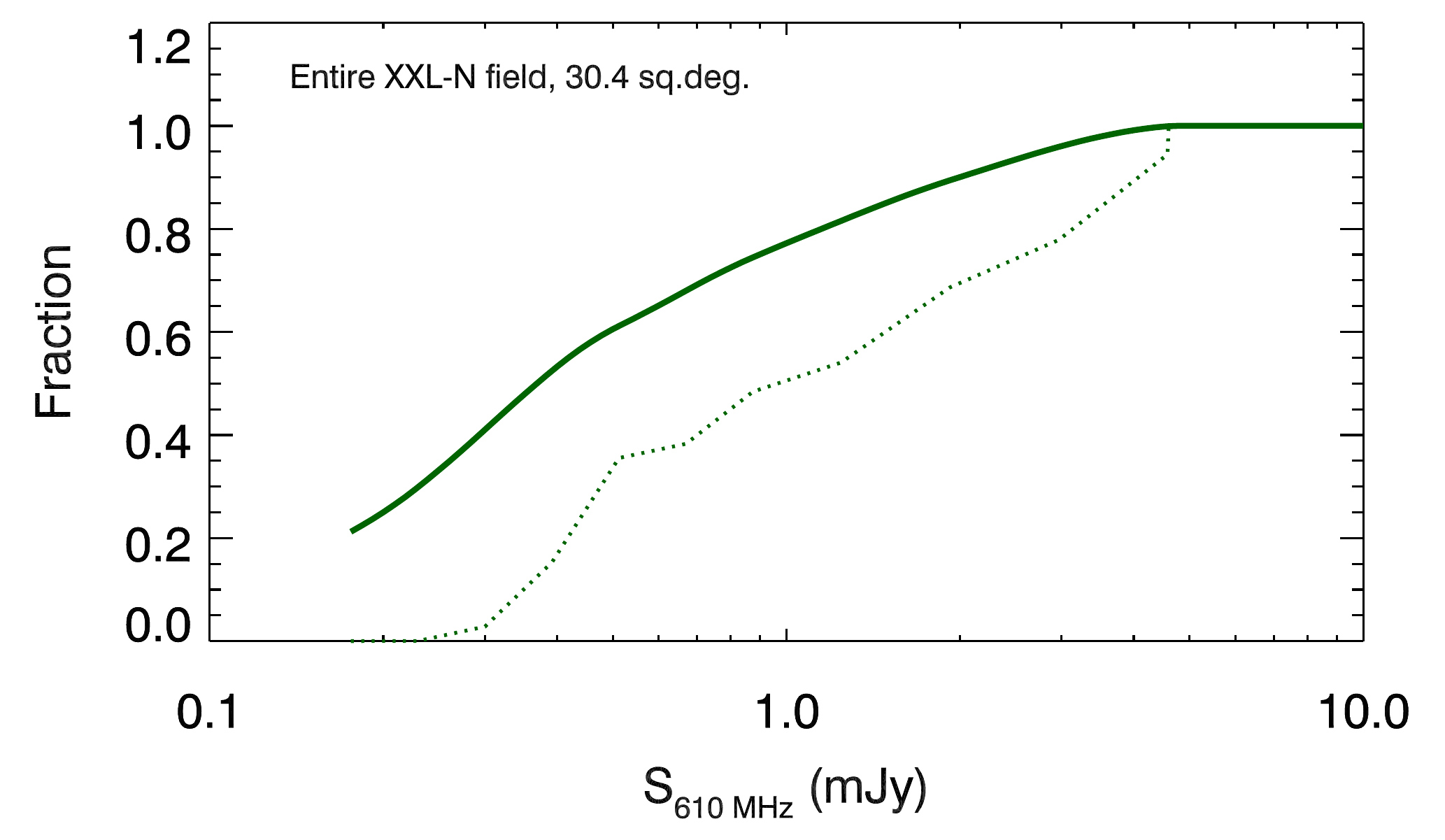}
\protect\protect\caption{
Same as \f{fig:compl2} \ but for the full 30.4 square degree observed GMRT-XXL-N area. 
\label{fig:compl2}
}
\end{figure}

\begin{table}
\begin{center}
\caption{Differential incompleteness measures, i.e. detection fraction as a function of 610~MHz flux density for the GMRT-XXL-N 610~MHz survey relative to the VLA-COSMOS 3 GHz Large Project source counts, and separated into two areas (inner and outer field) with different $1\sigma$ sensitivity limits. The estimated uncertainty of the detection fractions computed in this way is $\sim10\%$. Beyond the flux densities given here, we consider the survey complete (see text for details).}
\renewcommand{\arraystretch}{1.5}
\begin{tabular}[t]{c c | c c}
\hline
inner area & (9.6 deg$^2$)  & outer area  & (12.66 deg$^2$) \\
   \hline
 $S_\mathrm{610MHz}$ & detection &  $S_\mathrm{610MHz}$ & detection \\
  $[$mJy$]$ &  fraction &   $[$mJy$]$ &  fraction\\
  \hline
%   & flux [mJy] & detection fraction\\
%   & [mJy] & fraction \\
% from Mladen/complet_xxl_outer.tab1.24   &  0.09 &  0.30   &   0.07 \\
 1.92    &  0.45 & 0.39    &   0.34 \\
  2.96    &  0.65 & 0.51   &    0.75 \\
 4.58    &  0.94 & 0.67   &    0.80 \\
 & &  0.87   &    0.87 \\
  & &   1.24   &    0.87 \\
  & &    1.92   &    0.92 \\ 
  & &   2.96    &   0.96 \\
  & &   4.58   &    0.95 \\
    \hline
% from Mladen/complet_xxl_outer_cutedges_vla.tab
\end{tabular}
\label{tab:compl}
\end{center}
\end{table}

\section{Summary and conclusions\label{sec:summary}}

Based on a total of 192 hours of observations with the GMRT towards the XXL-N field we have presented the GMRT-XXL-N 610 MHz radio continuum survey. Our final mosaic encompasses a total area of 30.4 square degrees with a non-uniform rms, being $\sim200~\mu$Jy/beam in the inner area (9.6 square degrees) within the  XMM-LSS field, and $\sim45~\mu$Jy/beam in the outer area (12.66 square degrees). We have presented a catalogue of  5\,434 radio sources with S/N ratios down to $7\times$\,rms. Of these, 768 have been identified as components of  337 larger sources with complex radio morphologies, and flagged in the final catalogue. The astrometry, flux accuracy, false detection rate and completeness of the survey have been assessed and constrained. 

The derived 1.4 GHz radio source counts reach down to flux densities of $\sim150~\mu$Jy (corresponding to $\sim290~\mu$Jy at 610~MHz frequency assuming a spectral index of $\alpha=-0.7$, which is consistent with the average value derived for our sources at the bright end). Past studies have shown that the radio source population at these flux densities is dominated by  AGN, rather than star forming galaxies \citep[e.g.][]{wilman08,padovani15,smo08,smo17}. This makes the GMRT-XXL-N 610 MHz radio continuum survey, in combination with the XXL panchromatic data a valuable probe for studying the physical properties, environments, and cosmic evolution of radio AGN.

\section*{Acknowledgements}

This research was funded by the European Union's Seventh Framework programme under grant agreements 333654 (CIG, 'AGN feedback'). V.S., M.N. and J.D. acknowledge support from the European Union's Seventh Framework programme under grant agreement 337595 (ERC Starting Grant, 'CoSMass').  W.L.W. acknowledges support from the UK Science and Technology Facilities Council [ST/M001008/1]. We thank the staff of the GMRT who made these observations possible. GMRT is run by the National Centre for Radio Astrophysics of the Tata Institute of Fundamental Research. The Saclay group acknowledges long-term support from the Centre National d'Etudes Spatiales (CNES).
XXL is an international project based on an XMM Very Large \foreignlanguage{british}{P}rogramme
surveying two 25~square degree extragalactic fields at a depth of $\sim5\times10^{-15}$~erg~cm$^{-2}$~s$^{-1}$
in the {[}0.5-2{]}~keV band for point-like sources. The XXL website
is \url{http://irfu.cea.fr/xxl}. Multiband information and spectroscopic
follow-up of the X-ray sources are obtained through a number of survey
programmes, summarised at \url{http://xxlmultiwave.pbworks.com/}.

 \bibliographystyle{aa}
\bibliography{atca-xxl}

%\appendix
%dummy comment inserted by tex2lyx to ensure that this paragraph is not empty%dummy comment inserted by tex2lyx to ensure that this paragraph is not empty%dummy comment inserted by tex2lyx to ensure that this paragraph is not empty%dummy comment inserted by tex2lyx to ensure that this paragraph is not empty%dummy comment inserted by tex2lyx to ensure that this paragraph is not empty%dummy comment inserted by tex2lyx to ensure that this paragraph is not empty\bsp

%\label{lastpage} 
\end{document}